\documentclass[]{aastex631}
\usepackage{amsmath}
\usepackage{tipa}
\usepackage{mathtools}
\usepackage{makecell}
\usepackage{graphicx}
\usepackage{color}
\usepackage{nicematrix}
\usepackage{colortbl}
\usepackage{multirow}
\usepackage{capt-of}

\DeclareRobustCommand{\okina}{%
  \raisebox{\dimexpr\fontcharht\font`A-\height}{%
    \scalebox{0.8}{`}%
  }%
}

\newcommand{\oneI}{1I/\okina Oumuamua}
\newcommand{\twoI}{2I/Borisov}

\newcommand{\OO}{\={O}tautahi-Oxford}

\newcommand{\monthyeardate}{\ifcase \month \or January\or February\or March\or %
April\or May \or June\or July\or August\or September\or October\or November\or %
December\fi~\number \year} 

\newcommand{\rh}{\ensuremath{r_{\mathrm{h}}}}

\DeclarePairedDelimiter\abs{\lvert}{\rvert}%
\makeatletter
\let\oldabs\abs
\def\abs{\@ifstar{\oldabs}{\oldabs*}}

\submitjournal{PSJ}

\shorttitle{}
\shortauthors{Dorsey, Hopkins, Bannister et al.}

\begin{document}

\title{The visibility of the \OO{} interstellar object population model in LSST}

\correspondingauthor{Rosemary C. Dorsey}
\email{rosemary.dorsey@pg.canterbury.ac.nz}
\author[0000-0002-8910-1021]{Rosemary C. Dorsey}
\affiliation{School of Physical and Chemical Sciences --- Te Kura Mat\={u}, University of Canterbury, Private Bag 4800, Christchurch 8140, New Zealand}

\author[0000-0001-6314-873X]{Matthew J. Hopkins}
\affiliation{School of Physical and Chemical Sciences --- Te Kura Mat\={u}, University of Canterbury, Private Bag 4800, Christchurch 8140, New Zealand}
\affiliation{Department of Physics, University of Oxford, Denys Wilkinson Building, Keble Road, Oxford, OX1 3RH, UK}

\author[0000-0003-3257-4490]{Michele T. Bannister}
\affiliation{School of Physical and Chemical Sciences --- Te Kura Mat\={u}, University of Canterbury, Private Bag 4800, Christchurch 8140, New Zealand}

\author[0000-0001-5368-386X]{Samantha M. Lawler}
\affiliation{Campion College and the Department of Physics, University of Regina, Regina, SK S4S 0A2, Canada}

\author[0000-0001-5578-359X]{Chris Lintott}
\affiliation{Department of Physics, University of Oxford, Denys Wilkinson Building, Keble Road, Oxford, OX1 3RH, UK}

\author[0000-0002-6722-0994]{Alex H. Parker}
\affiliation{SETI Institute, Mountain View, CA, 94043, USA}

\author[0000-0002-1975-4449]{John C. Forbes}
\affiliation{School of Physical and Chemical Sciences --- Te Kura Mat\={u}, University of Canterbury, Private Bag 4800, Christchurch 8140, New Zealand}

\begin{abstract}

With a new probabilistic technique for sampling interstellar object (ISO) orbits with high efficiency, we assess the observability of ISOs under a realistic cadence for the upcoming Vera Rubin Observatory's Legacy Survey of Space and Time (LSST).
Using the \OO{} population model, we show that there will be complex on-sky structure in the pattern of direction and velocity revealed by the detected ISO population, with the expected enhanced northern flux complicating efforts to derive population parameters from the LSST's predominately southern footprint.
For luminosity functions with slopes of $2.5\leq q_s\leq 4.0$, the most discoverable ISOs have $H_r\simeq 14.6\text{--}20.7$; for previously estimated spatial densities, between 6 and 51 total ISOs are expected.
The slope of the luminosity function of ISOs will be relatively quickly constrained.
Discoveries are evenly split around their perihelion passage and are biased to lower velocities.
After their discovery by LSST, it will be rare for ISOs to be visible for less than a month; most will have $m_r \leq 23$ for months, and the window for spectroscopic characterization could be as long as two years.
These probabilistic assessments are robust against model or spatial density refinements that change the absolute numbers of ISO discoveries.
\end{abstract}

\keywords{Small Solar System bodies (1469), Interstellar objects (52), Interdisciplinary astronomy (804)}

\section{Introduction}
\label{sec:intro}

Interstellar objects (ISOs) provide insight into Galactic processes, from the properties of the protoplanetary disks that generate these planetesimals to star-formation histories and dynamics in the Galactic potential \citep{Fitzsimmons_2023}.
The dynamical processes operating within the Milky Way lead to a clumpy stellar velocity distribution in the Solar neighbourhood: a well-established set of moving groups and branches, recently made visible in exquisite resolution by \textit{Gaia} \citep{GaiaCollaboration_2023}.
As the same dynamical processes also act on the population of ISOs as they orbit in the Galaxy, \citet{Hopkins_2025} inferred that the ISO velocity distribution in the Solar neighbourhood has comparable substructure.
However, unlike stars, ISOs are too small and faint to detect while in the reaches between stars.
The substructure of the ISO population only becomes directly detectable from those objects that pass into the tiny part of the Solar neighbourhood that is the Solar System's observable volume.

Wide-field Solar System surveys are required to constrain the ISO number density and velocities in the observable volume, as highlighted by \citet{Francis_2005, Cook_2016} and \citet{Engelhardt_2017} even before the discovery of \oneI{}.
Only two ISOs are yet known: \oneI{} \citep{Meech_2017} and \twoI{} \citep{MPEC_2I}. 
Despite wide-field surveys such as Pan-STARRS, ATLAS and ZTF continuing to provide large areas of sky coverage to $m_r \sim 21.0\text{--}21.7$, obtaining a level of non-detection limits on the population, no further ISOs have been detected as of \monthyeardate.
Deeper surveys are needed.
In particular, the upcoming Vera C. Rubin Observatory's Legacy Survey of Space and Time (LSST) will provide routine coverage to $m_r \sim 24.0$ \citep{Bianco_2022b}: a substantial expansion of the observable volume.

Understanding the observed ISO sample requires a robust and data-driven population model.
Previous models of the Galactic interstellar object population have assumed a Gaussian velocity distribution for ISOs in the Solar neighbourhood
\citep{Whipple_1975, McGlynn_1989, Sen_1993, Engelhardt_2017, Meech_2017, Marceta_2023b}.
These assumptions have been used in previous assessments of the capabilities of the LSST for ISO detection and discovery \citep{Moro-Martin_2009, Cook_2016, Marceta_2023b}.
However, the ISO velocity distribution the Solar System can expect to encounter is highly non-Gaussian and clumpy \citep[as inferred from \textit{Gaia} by][]{Hopkins_2025}, so modelling needs to consider this more realistic population.

The physical properties of the local ISO population will be an integrated contribution of their source systems \citep{Moro-Martin_2018, Pfalzner_2019}, and could test if planetesimal processes are similar across the Galaxy. 
While we infer ISO age and metallicity from the \OO{} model, the distribution of ISO sizes depend on processes that are currently unconstrained.
Each system produces planetesimals, which then physically evolve, for instance under a collisional cascade, to develop a size-frequency distribution (SFD) particular to that sub-population within a system. 
Planetesimals may be unbound to join their system's ISO contribution in its tidal stream \citep{Forbes_2024} in any state from minimally to extensively evolved.
The population then sampled by the Solar System's observable volume could therefore range from the highly unlikely case of complete contribution from a single stream, to the more-probable case that each ISO comes from the stream of a different star system \citep{Forbes_2024}.
Planetary formation theory can provide a set of predictions; for example, planetesimals produced by the streaming instability will follow a comparatively shallow SFD \citep{Simon_2016}, but their number and mass contributed from a given stellar system depends on the stellar type and planetary architecture \citep[][and references therein]{Moro-Martin_2009}.
Observations, however, contribute an informative prior for steeper distributions; collisionally-evolved debris disks have an SFD that is only constrained at dust sizes \citep{Matthews_2014}, and the Solar System small-body size distributions \citep{Lambrechts_2016, Moro-Martin_2018, Bottke_2023} overlap with the two known ISO sizes.
Therefore, a range of ISO luminosity functions (and hence SFDs) must be considered.

Unlike most bound populations of Solar System small bodies, ISOs transition across visibility regimes and their survey observability is impacted by different biases.
Trailing losses may significantly reduce visibility of ISOs due to their potential high excess velocities \citep{Marceta_2023a}.
This typically occurs in the near-Earth space (similar to near-Earth asteroids) where ISOs exhibit large on-sky rates of motion.
At larger heliocentric distances, infalling ISOs instead have characteristics akin to trans-Neptunian discovery biases; ISO surveys must have suitable detection efficiencies throughout the heliocentric range.
This behaviour is somewhat shared by Solar System comets, whose detection \citep[typically at small distances;][]{Krolikowska_2019} is aided by extended comae. 
However, while long-period comets are isotropic in orbital distribution to first order, the Galactic motion of the Solar System means inbound ISOs pour preferentially across the sky from the Solar apex \citep{McGlynn_1989,Seligman_2018}, with an on-sky distribution affected by the local three-dimensional velocity distribution.

A survey with well-understood parameters is also key to the survey simulation of ISOs.
Historically, the complex observational properties of the LSST were often broadly approximated in such modelling; this was a natural outcome of the evolving creation of the cadence of the LSST, which eventually became fully community-generated \citep{Bianco_2022}.
These parameters are now converging: in particular, precise cadence modelling now exists.
\citet{Schwamb_2023} provided community assessment of its effects on a range of other Solar System populations, but only qualitatively approximated ISO requirements.
With first light for Rubin and the start of the LSST now approaching, it is timely to address LSST's capabilities for ISO discoveries in detail.

Here we explore the visibility of ISOs of the \OO{} ISO population model \citep{Lintott_2022,Hopkins_2023,Hopkins_2025,Forbes_2024} within the exact cadence and seeing conditions of the expected LSST. 
We provide probabilistic predictions of the ISO observability and discovery in LSST, with transparency to our model assumptions.
We also distinguish between multiple effects on observability to provide more generalised outcomes that demonstrate any survey's biases or sensitivities to ISOs.
We account for a range of luminosity functions in our observability modelling. 
We focus on inactive ISOs rather than those with comae, as the faintest limiting case of visibility.
Determining ISO physical properties will be a joint-facility effort: we predict the length of time ISOs will remain visible after their discovery by LSST, for further characterisation.

\section{ISO orbital model and orbit sampling}
\label{sec:sampling}

We use the \textit{Gaia}-derived model of the Solar-neighbourhood ISO population from \citet{Hopkins_2025}, the current version of the \OO{} model.
ISO tidal streams \citep{Forbes_2024} are not anticipated to substantially shift these kinematics.
In brief, a subsample of the \textit{Gaia} stars \citep{GaiaCollaboration_2016,GaiaCollaboration_2023} within 200~pc with measured 3D velocity, metallicity and age are debiased and reweighted to account for stellar death.
This \textit{sine morte}\footnote{We continue use of this from \citet{Hopkins_2023}: \textit{sine morte}, Latin for ``without death'', [\textprimstress si\textlengthmark ne \textprimstress m\textopeno rte] IPA pronunciation, or\\`seen-ay mort-ay'.} population, together with a protoplanetary disk model and associated assumptions, is used to predict the joint distribution in velocity, age and composition of ISOs in the Solar neighbourhood, far from the gravitational influence of the Sun.
Being calculated directly from stars in \textit{Gaia} DR3, it includes the complex structure of the local ISO velocity distribution, as well as the expected age and inferred water-mass fraction distribution of ISOs. 
 
ISOs provide additional challenges for survey simulation, as their orbits are unconfined to any given volume of space.
Integration of ISOs becomes more computationally expensive with increasing perihelia and the number of objects simulated.
A larger heliocentric orbital volume also needs to be modelled to account for gravitational focusing, where slow-moving ISOs are brought closer to the Solar System by the Sun's gravity, increasing the rate of ISOs with a low relative velocity passing through the inner Solar System.
Previous methods \citep[e.g.][]{Cook_2016,Engelhardt_2017, Seligman_2018, Hoover_2022} account for gravitational focusing numerically, starting by sampling a large number of ISO positions and velocities from the homogeneous background distribution, unperturbed and unfocused, from a volume around the Sun that is sufficiently large to make focusing effects negligible.
The resulting orbital state vectors were then integrated to produce a population of hyperbolic orbits.
Within this population, the ISOs of interest are those that enter a significantly smaller `observable sphere', a heliocentric volume of space in which an object is bright enough to be observed, with a radius dependent on the expected flux depth of a sky survey and the absolute magnitude \(H\) of the objects considered.
These `observable' ISOs represent a small fraction of the integrated orbits;
most initialised objects never pass close enough to the Sun and Earth to be observable due to their high perihelia and faint absolute magnitudes.
For example, out of the $\sim 2\times10^{9}$ objects initially synthesized by \citet{Engelhardt_2017}, only $10^6$ (or $<0.1\%$) were retained after applying orbital requirements $q\leq50$~au and $e>1$. 
This inefficient process becomes computationally expensive when simulating a large enough population of observable ISOs to perform meaningful inference. 

\cite{Marceta_2023b}'s ``probabilistic method", which samples the perturbed positions and velocities directly within an observable sphere while analytically accounting for gravitational focusing, is significantly more efficient.
However, this method only produces a sample of ISOs distributed around the Sun at a given moment (e.g. a single epoch or `snapshot').
Investigating the evolution of these objects within a time period, such as a Solar System survey, requires integrating each orbit.
The problem then becomes again that many sampled ISOs never enter their actual observable sphere, and the simulation must ``refresh" the population as new objects move in and out of their observable sphere during the survey duration.
The immediate solution would be to sample objects on an even larger sphere to encompass all objects that could potentially become observable, increasing in radius as a function of the survey duration and the maximum velocity of an ISO.
\citet{Marceta_2023a} used this approach, initializing a sphere with radius $\sim30$~au, and considering those ISOs that enter an \(\sim8\)~au radius\footnote{This corresponds to a maximum object diameter of 1~km.} observable volume --- but this only accounts for the motion of objects through the largest observable sphere within a single year of the upcoming LSST survey.
It also limits the maximum speed of the ISOs they consider to 22~au/yr (100~km/s), though this is not physically implausible \citep[see Fig.~3 in][]{Hopkins_2025}.
Modelling the full 10 years of LSST to account for ISO travel would require sampling a volume $10^3$ times greater; thus scaling this method to future longer surveys would result in the same problem of inefficiency as earlier. 

We take inspiration from the probabilistic method of \cite{Marceta_2023b} to construct a new ISO orbit initialization method.
Rather than initializing ISOs by sampling positions and velocities so that they can be propagated along their orbits, we sample their orbits directly using the \={O}tautahi-Oxford model velocity distribution.
The result is that every sampled ISO orbit enters its observable sphere at some time during a survey, accounting for the probabilistic distributions of orbital parameters, realistic non-Gaussian velocity distribution, and gravitational focusing.
This allows us to include non-zero lengths of time, and therefore avoid the need to sample ISOs in a larger volume of phase space than is realistically observable.
We simulate the full 10-year LSST survey with no upper limit on ISO speed and a much larger volume observable sphere, which we set to radius 100~au.
As in \cite{Marceta_2023b}, the orbits drawn are parameterized by their pre-encounter velocity, denoted with a heading in Galactic longitude and latitude \(l,b\) and a speed at infinity \(v_\infty\), as well as an impact parameter \(B\) and position angle \(\varphi\).
The distribution of orbits in these parameters depends only on the survey length \(T\) and the radius of the observable sphere \(r\), set by the minimum $H$ of the objects to be considered, and we derive this distribution here.
The flux of ISOs onto an orbit (\(l\), \(b\), \(v_\infty\), \(B\), \(\varphi\)) is given by \cite{Marceta_2023b} Eq.~11:
\begin{equation}
F_{l,b,v_\infty,B,\varphi} =\frac{\mathrm{d}N}{\mathrm{d}l\ \mathrm{d}b\ \mathrm{d}v_\infty\mathrm{d}B\ \mathrm{d}\varphi\ \mathrm{d}t} = n B v_\infty p_{l,b,v_\infty}
\end{equation}
where \(n\) is the ISO spatial number density and \(p_{l,b,v_\infty}\) is the velocity distribution of ISOs in the Solar neighbourhood, far from the gravitational influence of the Sun, for which we use the \OO{} model for the underlying velocity distribution (see Sec~2.3 and Fig. 6 of \cite{Hopkins_2025}). 

The number of objects on a given orbit that will be inside the observable sphere at any time during a survey of length \(T\) is equal to the sum of those in the sphere when the survey begins, plus the number that flow in over time \(T\). 
On the timescale of a survey, the flux of ISOs is constant \citep{PortegiesZwart_2021, Forbes_2024}, meaning this fluence is equal to:
\begin{equation}\label{eq:orbdist}
\frac{\mathrm{d}N}{\mathrm{d}l\ \mathrm{d}b\ \mathrm{d}v_\infty\mathrm{d}B\ \mathrm{d}\varphi} = \bigg(t_\mathrm{res}(v_\infty,B,r) +T\bigg) F_{l,b,v_\infty,B,\varphi} 
\end{equation}
for residence time \(t_\mathrm{res}(v_\infty,B,r)\), defined as the time spent by an object on a given orbit inside the observable sphere of radius $r$. 
An analytic expression for this and the integrals required for sampling this distribution are given in Appendix~\ref{sec:appendixOrbSamp}.
For \(T=0\), this reduces to the distribution of orbits of ISOs within a sphere of radius \(r\) at a single moment in time i.e. the distribution of \cite{Marceta_2023b}. 
For \(T\gg t_\mathrm{res}\), the ISOs already in the sphere when the survey begins are negligible compared to those that flow in over the survey, so the distribution is dominated by the refreshing population --- this is the volume-sampling weighted distribution used in \cite{Forbes_2019} and \cite{Hopkins_2025} (e.g. \(q<5\)~au in their Table 1 would correspond to this distribution's \(r=5\)~au).

Drawing \(l\), \(b\), \(v_\infty\), \(B\), and \(\varphi\) from the distribution in Eq.~\ref{eq:orbdist} produces a sample of orbits; we detail the method in Appendix~\ref{sec:appendixOrbSamp}.
To sample the positions of ISOs along these orbits, we choose the time of perihelion passage relative to the start of the survey \(\tau\), and for each orbit sample this from \(\text{Uniform}(-t_\mathrm{res}/2, \,\, t_\mathrm{res}/2+T)\). 
For example, an ISO with \(\tau=-t_\mathrm{res}/2\) had its closest approach to the Sun at time relative to the start of the survey at \(t=-t_\mathrm{res}/2\), and is outgoing on the edge of the observable sphere by the start of the survey at \(t=0\).
Similarly, an ISO with \(\tau=T+t_\mathrm{res}/2\) has its closest approach to the Sun at \(t=T+t_\mathrm{res}/2\), and only just enters the observable sphere at the end of the survey \(t=T\).

For computational simplicity, our method ignores some effects that will change the distribution of ISO orbits in a small way. 
Firstly, unlike \cite{Marceta_2023b} we ignore the negligible possibility that an ISO may not complete its Solar System passage due to an impact or close encounter with the Sun (similar to a sun-grazing comet).
For example, for $H_r=23$ objects within a $\sim2.16$~au observable sphere, only 1 in $\sim10^8$ objects of an isotropic ISO population would have orbits crossing the Solar Roche limit for asteroids \citep[$1.28R_{\mathrm{sun}}$ for an asteroidal density;][]{Gundlach_2012}.
This fraction decreases further for smaller values of $H_r$ as the observable sphere increases in radius.
In \S~\ref{sec:sfd} we confirm $\leq 0.02\%$ of discovered ISOs will be sungrazers.
Secondly, we assume all ISOs follow perfectly hyperbolic orbits about the Solar System barycentre, and thus ignore non-gravitational accelerations (i.e. cometary activity), deviations from Newtonian gravity due to general relativity, or close encounters with planets.

\subsection{Preparing the ISO Model for Observability Simulations}
\label{sec:shell_weighting}

Sampling ISO orbits within only one observable sphere over a multi-year-long survey is inefficient for understanding population characteristics, as the resulting population will be dominated by small objects on orbits too distant to be observable.
The maximum observable heliocentric distance in au, $\rh$, of an object at opposition is related to its absolute magnitude $H_r$ and apparent $r$-band magnitude $m_r$ by:
\begin{equation}\label{eq:r_max_obsSphere}
    \rh = \frac{1}{2}\left(1+\sqrt{1+4\times10^{\left(m_r - H_r\right)/5}} \right)\ \ (\mathrm{au})
\end{equation} 
For example, the maximum observable sphere to $m_r=25$ would be $\rh\sim100$~au for $H_r=5$, or $\rh\approx1.6$~au for $H_r=25$.
However, for an isotropic distribution of ISO orbits within a $\sim100$~au sphere, only 1 in $\sim244,000$ objects with $H_r=25$ (or 0.0004\%) would ever be observable during any part of their orbit.
This presents a large computational inefficiency.

To improve this, we divide the \(H_r\) range that we will consider (\(5\leq H_r \leq 23\), \S~\ref{sec:surveying}) into 1-mag intervals, and for each interval sample orbits passing through the observable sphere with radius \(\rh\) corresponding to the largest objects (minimum \(H_r\)) of each bin.
This increases the proportion of observable objects in each \(H_r\) interval with which to perform inference.
We draw 10,000 orbits in each 1-mag \(H_r\) interval for $5 \leq H_r \leq 13$ and 50,000 orbits in each 1-mag \(H_r\) interval for $13 \leq H_r \leq 23$.

To compare these samples from different \(H_r\) intervals as if they were drawn from the one observable sphere and one \(H_r\) interval they must be reweighed.
For each \(H_r\) interval this weighting is proportional to the number of objects within the \(H_r\) interval expected to enter the corresponding observable sphere.
The total number of objects entering an observational sphere of radius is given by the integral over \(\mathbf{v}_\infty\) of Eq.~\ref{eq:vdist}, which being non-analytic we approximate well with \(n\cdot\big(\frac{4}{3}\pi \rh^3 + \pi \rh^2 T(0.029\,\mathrm{au/day})\big)\) (cf. Fig.~\ref{fig:normapprox}).
The fraction of these objects within the corresponding size interval $H_r=\left[H_{\mathrm{min}}, H_{\mathrm{max}}\right]$ is proportional to $ \left(10^{\alpha H_{\mathrm{max}}} - 10^{\alpha H_{\mathrm{min}}}\right)$, for a given single slope absolute magnitude distribution
$dN/dH \propto 10^{\alpha H}$.
Thus, the total weight $w$ given to each \(H_r\) interval is given by:
\begin{equation}\label{eq:total_weighting}
   w \propto \left(10^{\alpha H_{\mathrm{max}}} - 10^{\alpha H_{\mathrm{min}}}\right) \left(\frac{4}{3}\pi {\rh}^3 + \pi {\rh}^2 T\left(0.029\,\mathrm{au/d}\right) \right)
\end{equation}

\section{Simulating ISO observability within the LSST}
\label{sec:surveying}

While we aim to identify common concerns for ISO visibility in sky surveys with our model, the primary survey we consider here is Rubin's LSST. 
This 10-year survey with the 8.4 m Simonyi Survey Telescope is expected to reach a single-image depth of $m_r \sim 24.0$ across some 18,000 square degrees of predominantly Southern sky \citep{LSSTScienceCollaboration_2009, Bianco_2022b}.
For our assessment of the visibility of ISOs, we use the LSST cadence simulation \texttt{baseline\_v3.3\_10yrs} \citep{LSSTv3.3, Naghib_2019}, the most realistic available during this work; the small changes since\footnote{As of the Phase 3 Recommendation: \url{https://pstn-056.lsst.io/}} affect neither the footprint nor filters of primary ISO detectability.
Our approach is to consider and identify the major influences on the probabilistic observability of our probabilistic ISO velocity model.

We construct a set of nomenclature:
\begin{itemize}
    \item \textbf{visited} --- a probabilistic ISO orbit is within an LSST field of view (FOV);
    \item \textbf{observed} --- a probabilistic ISO orbit with physical properties applied would be an $\geq 5\sigma$ source in an LSST image, forming an observation;
    \item \textbf{discovered}\footnote{The formal LSST nomenclature uses `detect' for this term, but since we are discussing multiple aspects of detectability, we use `discovered' for clarity.} --- a minimum set of LSST observations exist for an observed ISO that meets particular discovery requirements;
    \item \textbf{arc length} --- the period between the first and last observations of an ISO by LSST;
    \item \textbf{characterisation window} --- the period that an ISO has brightness $m_r \leq m_\mathrm{interest}$ after it is discovered. 
\end{itemize}

We first assess whether our ISOs are present within the FOV of LSST at the time of each observation in \texttt{baseline\_v3.3\_10yrs}.
For this, we evaluate the position of the ISO along its orbit at the time of the pointing \citep[using the method of][]{Farnocchia_2013}, and check if the ISO lies within a cone of opening angle \(\theta_\mathrm{FOV}=1.75\degr\) pointing from the Earth's position at that time.
This assumes a simplistic circular FOV, and that objects remain within the FOV for the whole exposure.
For computational efficiency, we skip some pointings guaranteed not to be visited: for orbits with perihelion \(q>1\)~au, these are pointings perpendicular to the plane of the orbit and pointings directly away from perihelion, extending to any pointings within an angle of \(\arccos(1~\mathrm{au}/q) - \theta_\mathrm{FOV}\) of the perpendicular to the orbit, and any pointings within an angle of \(\arccos(1/e) - \theta_\mathrm{FOV}\) of the direction away from perihelion. 
If a given ISO meets these criteria for at least one pointing, it is \textbf{visited}.

Once an ISO's orbital geometry has meant it is visited, we apply physical properties to determine whether it is \textbf{observed}.
An object's apparent magnitude in an LSST filter band at any given time is described by:
\begin{equation}\label{eq:m_filter}
    m_{\mathrm{filter}} = H_r
    + \underbrace{%
    \vphantom{\left( 1 + \frac{ax^2}{1+bx} \right)}
    5\log_{10}(\rh\Delta)}_{\Delta m_{\mathrm{dist}}}
    + \underbrace{%
    \vphantom{\left( 1 + \frac{ax^2}{1+bx} \right)}
    \abs{2.5\log_{10}(\Phi)}}_{\Delta m_{\mathrm{phase}}}
    + \underbrace{%
    \vphantom{\left( 1 + \frac{ax^2}{1+bx} \right)}
    \abs{1.25\log_{10} \left( 1 + \frac{ax^2}{1+bx} \right)}}_{\Delta m_{\mathrm{trail}}}
    + \underbrace{%
    \vphantom{\left( 1 + \frac{ax^2}{1+bx} \right)}
    \left(m_{\mathrm{filter}}-m_r\right)}_{\mathrm{colour}}
\end{equation}
\begin{equation}
    x = \frac{\mu T_{\mathrm{exp}}}{24\theta}
\end{equation}

where $H_r$ is the absolute magnitude (we use $r$-band for consistency), \rh~is the heliocentric distance in au, $\Delta$ is the geocentric distance in au, $\Phi$ is the phase function evaluated at phase angle $\alpha$,
and $a$ and $b$ are trailing loss coefficients with values given in Table~\ref{tab:trailing_loss_coefficients}.
The variable $x$ is a function of the on-sky rate of motion $\mu$ in degrees per day, the image exposure time $T_{\mathrm{exp}}$ in seconds, and the median seeing $\theta$ in arcseconds.
The variables $T_{\mathrm{exp}}$ and $\theta$ are provided by the LSST cadence simulation for each observation.
We factorise Eq.~\ref{eq:m_filter} into five independent terms, describing the effect of morphologic ($H_r$), geometric ($\Delta m_{\mathrm{dist}}$, $\Delta m_{\mathrm{phase}}$), kinematic ($\Delta m_{\mathrm{trail}}$) and compositional (colour) characteristics of ISOs on their observed apparent magnitude.
The inverse-square distance effects on brightness ($\Delta m_{\mathrm{dist}}$) are straightforward, but the other terms have additional complexities.

\begin{table}[b]
    \centering
    \begin{tabular}{l|c|c}
         & \multicolumn{2}{c}{Coefficient} \\
         Trailing Loss Type & $a$ & $b$ \\ \hline
         SNR only & 0.67 & 1.16 \\
         SNR \& Detection & 0.42 & 0.00 \\
    \end{tabular}
    \caption{Values for the trailing loss coefficients in Eq.~\ref{eq:m_filter}.}
    \label{tab:trailing_loss_coefficients}
\end{table}

The absolute magnitude of an individual object is a direct result of its physical characteristics, including size (diameter~$D$ in~km), dimensions (axis ratio~$a$:$b$:$c$) and surface reflectivity (albedo~$p$, which is dependent on composition).
These properties are influenced by disk formation location, planetesimal growth efficiency, collisional evolution, and thermal processing.
Their impact on the characteristics of individual ISOs and their overall Galactic population is currently unconstrained.
While well explored for Solar System small-body populations, with only two known ISOs \citep[e.g.][]{Jewitt_2023} the parameter space is open.
For simplicity, we consider a single-slope power law and test a range of slope values $q_s=[2.5, 3.0, 3.5, 4.0]$ that bracket the streaming instability-like SFD \citep[$q_s=2.8$,][]{Simon_2016}.
These cover a range of values from the absolute magnitude distributions of different Solar System small-body populations, including asteroids \citep{Bottke_2005, Alvarez-Candal_2006}, Kuiper Belt objects \citep{Petit_2023}, as well as the predicted primordial planetesimal disk \citep{Bottke_2023}.
The minimum value for $q_s$ is set to the critical value below which larger objects are more probable discoveries than smaller objects; see Appendix~\ref{appShallowSlope} for a detailed derivation.
We include the streaming instability size distribution $q_s=2.8$ in our analysis for several reasons.
Firstly, the majority of circumstellar material is ejected from its planetary system during the early stages of planet formation and evolution \citep{Fernandez_1978, Levison_1997, Raymond_2020, Fitzsimmons_2023} and therefore has potentially undergone minimal processing via collisional or thermal effects.
Additionally, our ISO model (\S~\ref{sec:sampling}) assumes that the age of an ISO is directly correlated to the age of the star that produced it.

We assess observability for objects with absolute magnitudes $5 \leq H_r \leq 25$, corresponding to ISOs of size $0.07\lesssim D \lesssim 600$~km for an assumed asteroidal albedo $p_r=0.05$ \citep{Licandro_2016, Masiero_2021}.
For context, the upper limit corresponds to the minimum diameter at which internal differentiation could be expected, also used for an object to be classified as a rocky dwarf planet \citep{Lineweaver_2010}, while the lower limit encompasses the sizes of the two known ISOs, radius of $\sim100$~m for \oneI{} \citep{`OumuamuaISSITeam_2019} and $\sim0.5$~km for \twoI{} \citep{Jewitt_2020, Hui_2020}, as well as the expected LSST $r$-band single image 5$\sigma$ limiting magnitude \citep[$m_{5}\sim24.0$;][]{Bianco_2022b}.

Other parameters that influence the absolute magnitude distribution of ISOs include size~$D$, albedo $p_r$ shape $a$:$b$:$c$, and light curve amplitude.
While the size and shape of \oneI{} are reasonably well-constrained by its light curve \citep{`OumuamuaISSITeam_2019}, their values for the nucleus of the cometary ISO \twoI{} are only inferrable from the physical extent of its coma and the non-gravitational perturbations to its orbit, with both methods requiring assumptions for the nucleus composition, albedo, density, and structure.
The albedos of the known ISOs are estimated to be in the range $p \sim 10^{-2}\text{--}10^{-1}$ \citep{Bannister_2017, Trilling_2017, Jewitt_2019}; however, the true values are not well-constrained.
Additionally, there exists a correlation between size and shape among Solar System objects with $D>1$~km \citep[e.g. main-belt asteroids][]{Mommert_2018}; however for small objects $D<200$~m this relationship remains uncharacterized \citep{Mainzer_2023, McNeill_2019, Thirouin_2018}.
Thus, while the extreme light curve variability of \oneI{} undoubtedly influenced its observability \citep{Levine_2023} (see also \S~\ref{sec:unobservability}), it is unwise to extrapolate the ISO shape distribution from a single known value.
Therefore, we leave the parameters of albedo, shape, and light curve amplitude independent in our modelling, which only impact ISO observability via their relation to absolute magnitude.

For the geometric effect $\Delta m_{\mathrm{phase}}$, we calculate the phase function $\Phi=\Phi(\alpha)$ assuming a spherical solid body of uniform albedo \citep[Eq.~36 in][]{Bowell_1989}.
The phase functions of minor planets are well-explored \citep{Colazo_2021, Dobson_2023, Alvarez-Candal_2024, Robinson_2024}; however, defining relationships that accurately model the opposition surge for all populations remains a work in progress.
Minimal phase angle coverage of both known ISOs during their Solar System passages resulted in little information on their phase functions \citep{Fraser_2018, Lu_2019, Opitom_2019, deLeon_2020}.
In particular, the orientation and timing of their trajectories relative to the Earth meant that neither was observed near $\alpha=0^\circ$, where opposition surge significantly impacts the phase function.
In addition to the sparsely sampled phase angles, phase function is also dependent on colour, albedo, shape or axis ratio (acute for \oneI{} cf. \citet{Mashchenko_2019}, but less pronounced for other small-body populations, and not known for \twoI{}) and cometary activity (seen strongly for \twoI{} \citep[e.g.][]{Bannister_2020}, but only at upper limits for \oneI{} \citep{Micheli:2018}).
Since ISOs are expected to be observed across a wide range of phase angles in LSST, we assume a simplistic phase function model until their phase behaviour is better characterised.

We next consider the kinematic losses.
These are directly tied to the velocity distribution of our input ISO model.
Trailing loss $\Delta m_{\mathrm{trail}}$ is the reduction in the observed apparent magnitude of an object due to its on-sky rate of motion relative to the telescope tracking \citep{Jones_2018}.
This appears in photometric images as a non-stellar point spread function (PSF) distribution of flux, or `trail'.
Trailing loss has two effects on the detectability of an object in a given exposure.
The first is an inherent reduction in the signal-to-noise ratio (SNR), due to the signal spreading over a larger area of background noise pixels than a stationary source.
The second effect, detection loss, is due to the non-optimal use of a circular PSF to identify new sources, which does not fully encapsulate the total flux of the target.
Unlike SNR trailing loss, it is possible to prevent detection loss by adapting the shape of the PSF used in the source detection software to account for large on-sky motion.
The primary Rubin LSST source detection pipeline does not use non-stellar PSF filters \citep{Schwamb_2023}\footnote{The subsequent LSST Solar System Processing uses trailed PSFs to link observations, however.}.
Detection losses must therefore be considered when modelling the detectability of ISOs as sources in LSST visits.
Trailing loss is typically negligible for objects in the outer Solar System, which have very slow on-sky velocities on the order of $\sim1 \arcsec$~hr$^{-1}$, but needs to be considered from the orbit of Mars inward.
The observability of objects with large on-sky velocities on the order of $\sim0.1\text{-}10\degr$~d$^{-1}$ (such as ISOs at $\rh<$1-2~au) will thus be impacted by trailing losses. 

While Solar System small bodies display a wide range of colours in the optical-NIR, there is only one surface colour measurement for the known ISO population.
\oneI{} was observed to be red, similar to Jupiter Trojans, Jupiter family comets and long-period comets \citep{Jewitt_2017, Bannister_2017}, while the nucleus of \twoI{} was obscured by its coma for its entire observable passage through the Solar System; hence the colour distribution of the intrinsic ISO population is unconstrained.
We tested the sensitivity of ISO discovery in LSST to the LSSTCam $ugrizy$ distributions for Solar, S-type, C-type, D-type and CCKBO asteroidal surface types \citep{Willmer_2018,Schwamb_2023}.
We found no significant correlations between colour and the likelihood of discovery, and the orbital and physical properties of the discovered ISOs.
Hence, we adopt a solar spectral type for ISOs in this work.

We do not directly model satellite streaks and their complex impact on source detection in LSST images \citep{Tyson_2020, Hu_2022}.
Instead, we performed a preliminary analysis of the influence of randomly dropped images on ISO discovery in LSST, to approximate the discovery losses due to various phenomena that we do not model in this work (e.g. difference imaging artifacts, saturated stars etc).
Observations were removed with 5\% probability near the Galactic plane and 1\% probability elsewhere.
Although discovery rates decreased by $\sim2\text{--}5\%$ (increasing with slope $q_s$), no significant biases in the resulting orbital or physical parameters of the discovered ISO population were identified.

The probability of an object being observed in a given image is then given by:
\begin{equation}
    P = \left[ 1 + \exp \left(\frac{m-m_5}{\sigma} \right) \right]^{-1}
\end{equation}
where $m$ is the object's apparent magnitude in the given filter band (any of $ugrizy$ for LSST), $m_5$ is the 5$\sigma$ limiting magnitude in the given filter band, and $\sigma$ is the width of the completeness falloff.
The LSST pipeline computes sources by subtraction of individual survey images against a template of each region of the sky that will be developed from the first year of survey operations.
We assume detection occurs within a single image with differencing from the template as per the LSST~SSP, i.e. there is no application of shift-and-stack for object detection to deepen the 5$\sigma$ limiting magnitude.
We assume $\sigma=0.12$, in accordance with \citet{Jones_2018}.
An object's detection probability $P$ is calculated for each visit in an LSST image.
These probabilities are randomly sampled to produce a set of visits where the object is \textbf{observed}.

\subsection{Effects on Observability} 
\label{sec:detectability}

To understand the effect of different brightness variation mechanisms on the observational probability of ISOs, we first consider a preliminary population of 10,000 ISO orbits initialised within a single heliocentric sphere of radius 8~au.
We assign a constant absolute magnitude $H_r=17.4$ to each orbit; for an assumed albedo, e.g. $p_r=0.05$, this corresponds to a diameter of $D\sim2$~km.
We only consider images acquired in the first year of LSST.

\begin{figure}[]
    \centering
    \includegraphics[width=\linewidth]{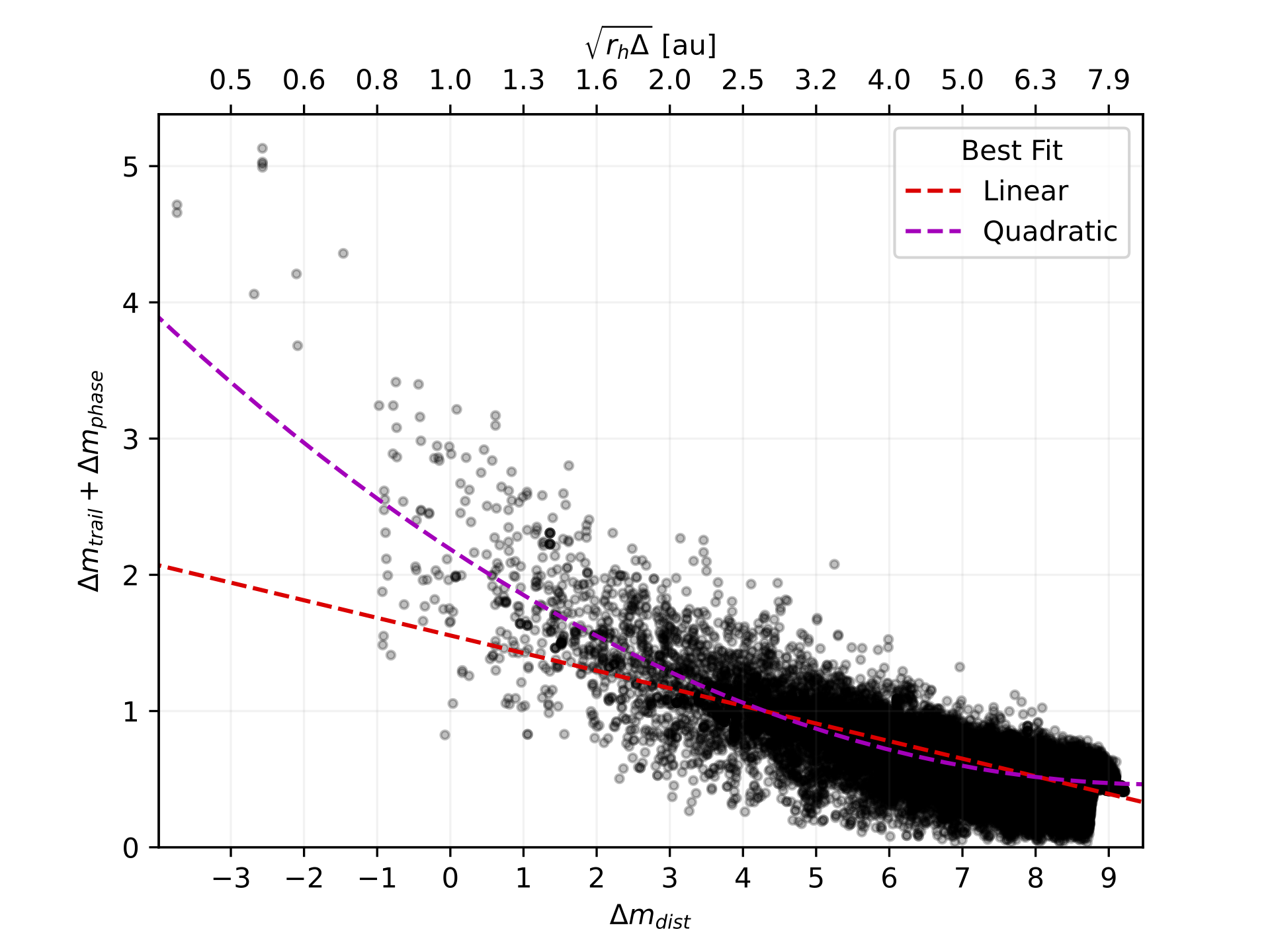}
    \caption{Combined effect of phase function and trailing loss as a function of $\Delta m_{\mathrm{dist}}$. 
    Best fits using a first-order and second-order polynomial are provided for qualitative reference only. 
    The trend is almost linear but diverges near the inner Solar System ($\Delta m_{\mathrm{dist}} \lesssim 4.0$). 
    Note that this relationship is independent of morphological effects such as absolute magnitude ($H_r$) and colour.}
    \label{fig:trailPhaseVariability}
\end{figure}

We first consider the relationship between distance and the secondary apparent magnitude loss due to phase and trailing.
Unlike most other small-body populations, a hyperbolic ISO orbit can have any orientation, so observations can be made at any phase angle $\alpha \in [0,\pi]$ with potentially large on-sky velocities $\mathcal{O}({^\circ/d})$.
This means that a simple distance-based limiting magnitude systematically overestimates the maximum observable distance of an ISO.
Figure~\ref{fig:trailPhaseVariability} demonstrates how the combined effect of trailing loss and phase varies with distance, independently of an object's absolute magnitude.
The relationship appears linear for large $\Delta m_{\mathrm{dist}}$ and $\sqrt{\rh\Delta}$, but diverges for $\Delta m_{\mathrm{dist}} \leq 4$.
This means that objects at small heliocentric and geocentric distances will be disproportionally more affected by phase and trailing losses than objects at large distances.
At $\Delta m_{\mathrm{dist}} = 2.0$, the combined phase and trailing losses deviate by $\sim0.25$ from the linear approximation.
This corresponds to objects with a maximum observable heliocentric distance of $\rh \approx 2.16$~au, or $H_r \approx 23$.
We take this as an effective ``limiting absolute magnitude" for ISOs within LSST, and all further analysis in this work uses $H_r\leq23$.

\begin{figure}
    \centering
    \includegraphics[width=0.65\linewidth]{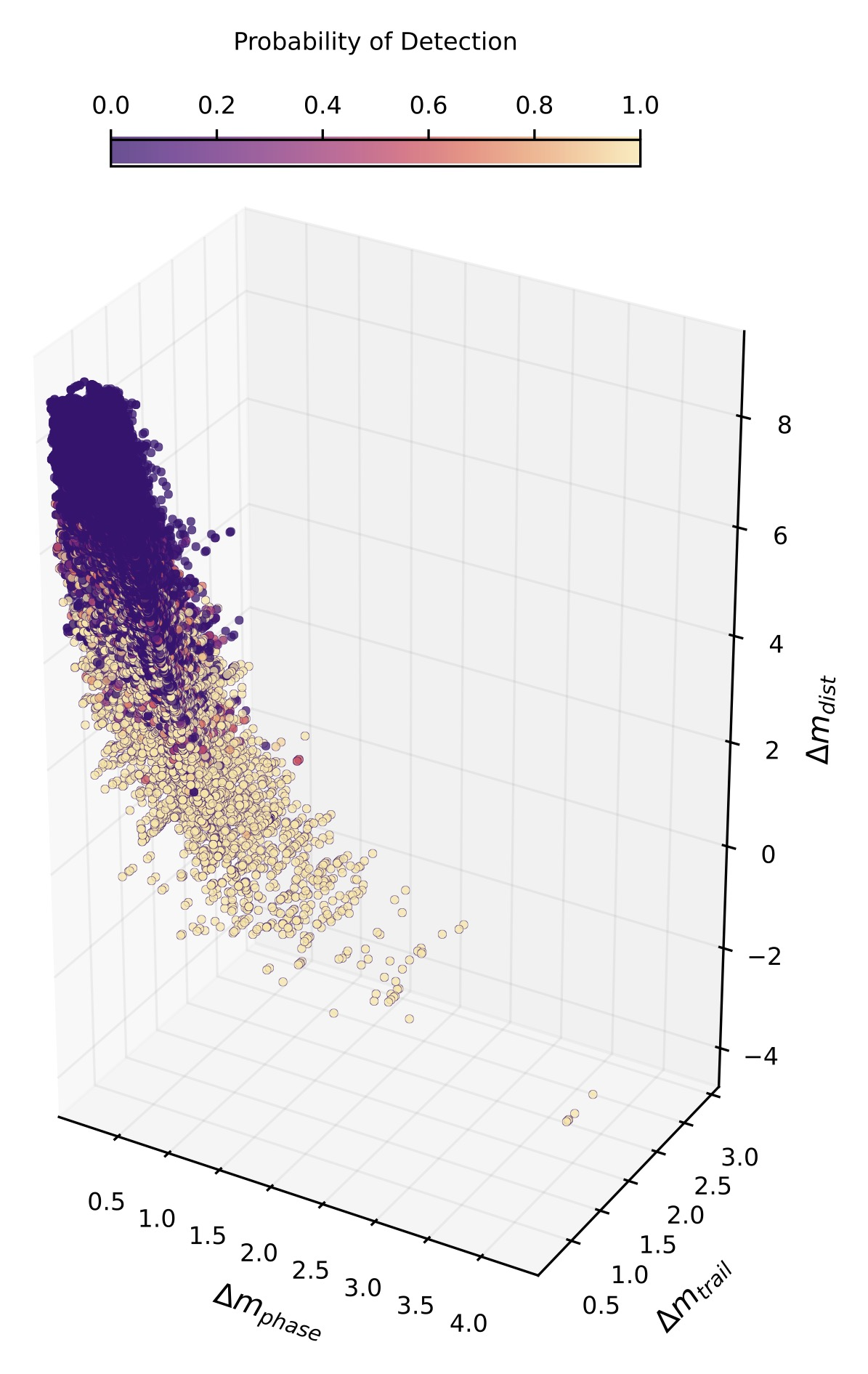}
    \caption{Influence of the geometric and kinematic observational biases on the detection probability of all instances where an ISO's orbital position is synchronised with the LSST cadence. The axes refer to the last three terms in Eq.~\ref{eq:m_filter}, corresponding to the change in apparent magnitude of an object due to distance ($\Delta m_{\mathrm{dist}}$), geometric viewing ($\Delta m_{\mathrm{phase}}$) and trailing loss ($\Delta m_{\mathrm{trail}}$). The nonlinearity in the transition from completely observable ($P=1$) to completely unobservable ($P=0$) ISOs is due to the seeing and limiting magnitude varying across all LSST images.}
    \label{fig:3d}
\end{figure}

We then consider the effect of the geometric viewing and kinematic losses on the probability that an ISO is observed in an LSST image.
Unsurprisingly, the dominant effect on ISO observability is from their distance at observation (Fig.~\ref{fig:3d}).
However, phase and trailing losses also affect observability.
For the same $\Delta m_{\mathrm{dist}}$, images taken at high phase angles with high on-sky motions are less likely to produce a successful ISO observation than images taken at low phase angles with low on-sky motions.
An additional interesting feature of Figure~\ref{fig:3d} is the granular nature of the transition in observational probability.
This is due to the variation in the $5\sigma$ limiting magnitude and seeing between individual LSST observations.
These variations model the real-world physical phenomena that will affect the LSST cadence (e.g. seasonal weather patterns, airmass variations etc), hence creating a complex relationship between the three variables in Figure~\ref{fig:3d}.

\subsection{Discoverability}\label{sec:detectability_definition}

We assess the discoverability of ISOs within the context of the LSST Solar System Processing \citep[hereafter referred to as LSST~SSP;][]{Myers_2013, Juric_2020, Claver_2024}.
The criteria for the discovery of a moving object is defined by the Observatory System Specifications requirement OSS-REQ-0159 \citep{Claver_2024}, which we list verbatim: an object is \textbf{discovered} by the LSST~SSP $\geq95\%$ of time when it is observed in the following pattern:
\begin{itemize}
    \item over 15 days, the object is observed on at least three separate nights, with at least one tracklet per night, where
    \item a tracklet is defined as two or more observations taken in an interval not longer than 90 minutes where an object's astrometric position has changed by $\geq3\sigma$ \citep[where $\sigma \approx 140$~mas is the positional astrometry uncertainty;][]{Jones_2018, Schwamb_2023}, and
    \item each observation being defined as a source registering above the 5-sigma threshold after convolution with the image’s PSF (as required by LSST source detection).
\end{itemize}

This criteria applies to the main survey, comprised of the `Wide-Fast-Deep' and the Deep Drilling Fields, but breaks down for the fraction of LSST observations performed during the low-elongation twilight survey every few nights.
This microsurvey targets small Solar System bodies that spend limited time within the main LSST footprint, including objects with fast on-sky motion rates (such as NEOs or ISOs) as well as inner Earth orbital populations \citep[e.g. the recently discovered `Ayl\'{o}'chaxnim asteroid population;][]{Bolin_2020, Bolin_2022}, which are only observable within the twilight survey.
Analysis by \citet{Schwamb_2023} showed that less stringent and more conventional object detection methods such as three or four observations within the same night could significantly improve the discovery completeness for such fast-moving objects.
Thus, in addition to the standard LSST~SSP detection criteria employed for the main survey, we assume that observing an object in four twilight visits within the same 90-minute interval is sufficient for its discovery.

\begin{figure}[]
    \centering
    \includegraphics[width=\textwidth]{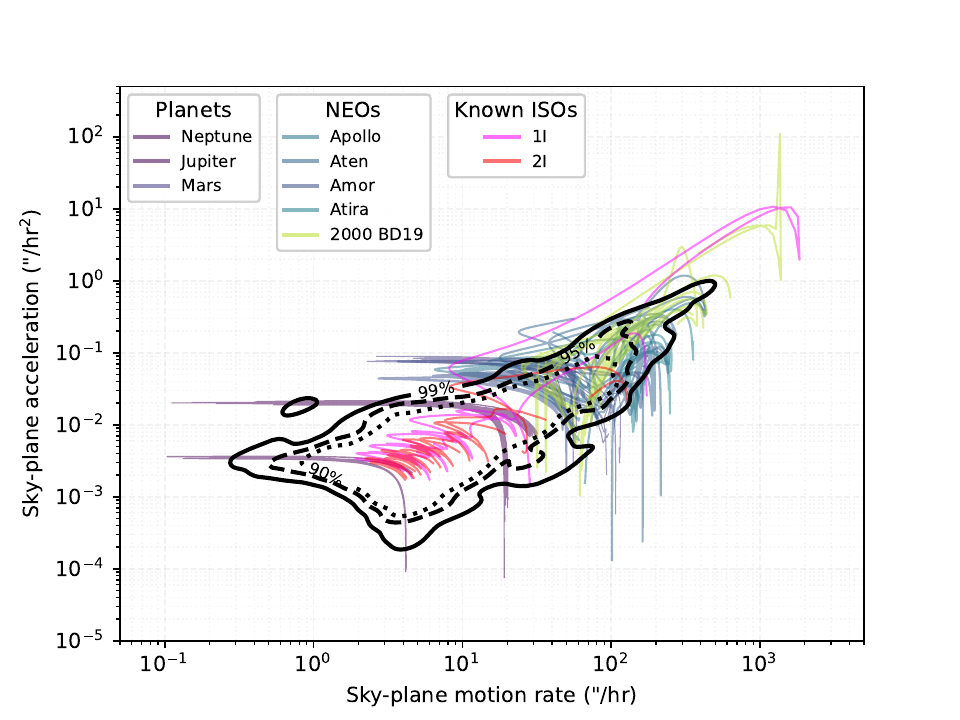}
    \caption{Expected on-sky rates of motion and acceleration for simulated ISOs with $q\leq5$~au within LSST visits (regardless of whether they are observed in an image) compared to a selection of bound Solar System objects and the two known ISOs. 
    The planets and NEOs were simulated over the ten-year duration of LSST using ephemerides from JPL Horizons, whereas \oneI{} and \twoI{} were simulated for the same duration but centred around their perihelion passage through the Solar System. 
    Objects with low perihelia appear toward the top right, while objects with high perihelia appear toward the bottom left.
    Note that an on-sky rate of motion of 1~${^\circ}$~d$^{-1}$ is equivalent to 150~\arcsec~hr$^{-1}$.
    }
    \label{fig:motion_rate}
\end{figure}

ISOs are expected to exhibit a large range of on-sky rates of motion as they pass through the Solar System, due to their high hyperbolic excess velocities \citep[cf. Fig. 3 in][]{Hopkins_2025}.
Although LSST will be capable of linking large trails \citep[$\gtrsim 1$~$^\circ$~d$^{-1}$;][]{Jones_2018}, the on-sky motion rate is typically conservatively constrained to $\sim0.5$~$^\circ$~d$^{-1}$ for computational efficiency.
This potentially raises concerns about the linking of ISO tracklets.
Figure~\ref{fig:motion_rate} explores the range of on-sky motion rates for each LSST visit to simulated ISOs with perihelia $q\leq5$~au.
In more than 95\% of visits, the ISO has on-sky velocity $\leq150$~\arcsec~hr$^{-1}$ (or 1~$^\circ$~d$^{-1}$), while $>99$\% of all visits to all simulated ISOs have on-sky velocities $\leq100$~\arcsec~hr$^{-1}$.
Additionally, in at least 99\% of visits, ISOs with $q\leq5$~au will have on-sky motion rates and accelerations consistent with bound Solar System objects (including outer Solar System planets and NEOs) and the two previously observed ISOs. 
Thus, we can expect only a fraction of LSST visits will be to ISOs exhibiting high on-sky motion rates, and these are restricted to orbits with perihelia $q\leq5$~au; in \S~\ref{sec:sfd}, we confirm with our simulations $\leq 0.05\%$ of discoveries will have on-sky motion faster than the LSST SSP threshold of 10~$^\circ$~d$^{-1}$ \citep{OMullane_2024}.
We therefore assume SSP's linking is consistently successful.

\section{Constraints on the ISO Size-Frequency Distribution}
\label{sec:sfd}

\begin{figure}
    \centering
    \includegraphics[width=\linewidth]{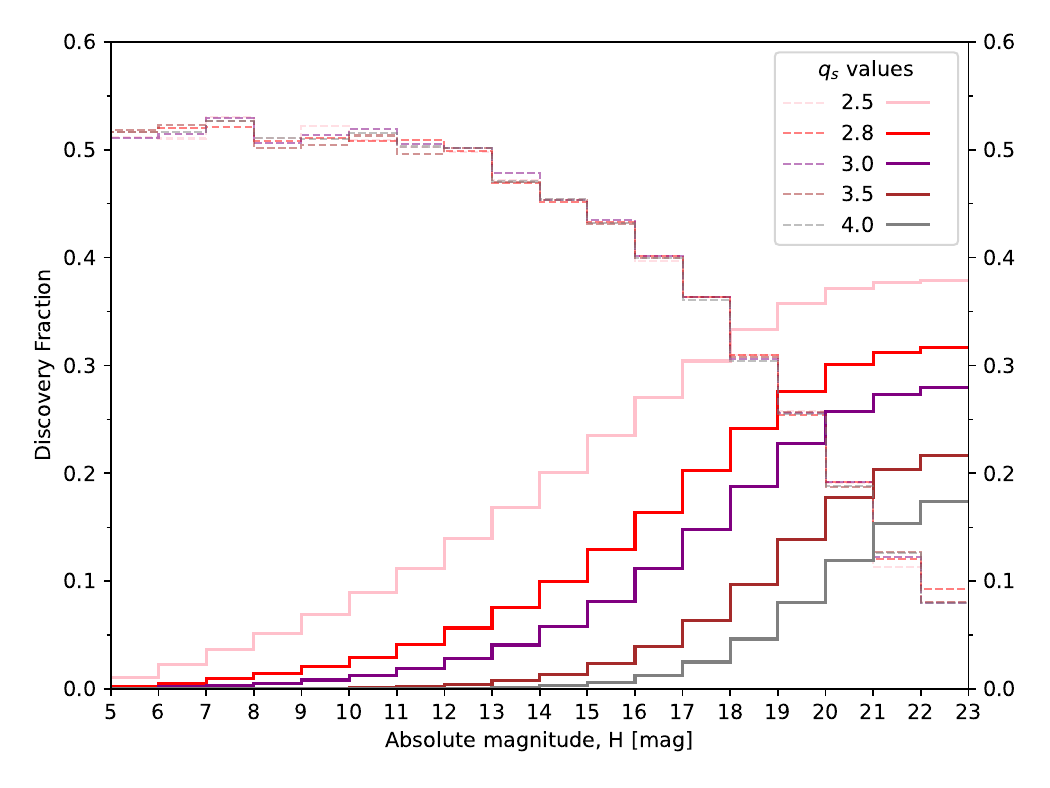}
    \caption{Discovery completeness functions for the absolute magnitude of observed ISOs in LSST per $H_r$ interval (dashed lines) and cumulatively (solid lines). For $H_r<13$, 50\% of observed ISOs are discovered; however, this discovery rate decreases for objects with $H_r\geq13$, where the lowest discovery rate is associated with the smallest objects ($<10\%$ for $H_r\geq22$). Regardless of the intrinsic absolute magnitude distribution, $<40\%$ of the ISOs observed by LSST were discovered.}
    \label{fig:discCompleteness}
\end{figure}

The discovery efficiency of LSST is significantly incomplete for ISOs, even within the inner Solar System.
Of the ISOs observed by LSST (i.e. objects with at least one visit), 35.2--62.2\% have $\geq6$~detections and 27.9--52.8\% have $\geq3$~tracklets in the main survey, with the steepest absolute magnitude slope $q_s=4.0$ population being the least visible.
The fraction of observed objects that result in LSST discoveries is shown in Figure~\ref{fig:discCompleteness}.
Even for the largest ISOs, observed objects are only discovered $\sim50\%$ of the time.
Discovery completeness decreases for smaller objects with $H_r\geq13$, reaching $<10\%$ for the smallest of these ($H_r\geq22$).
Across all $H_r$ values, $<40\%$ of the ISOs observed in at least one LSST image will be discovered, with the largest slope parameter $q_s=4.0$ model being the least discoverable population at $<20\%$.

\begin{figure}[]
    \centering
    \includegraphics[width=\linewidth]{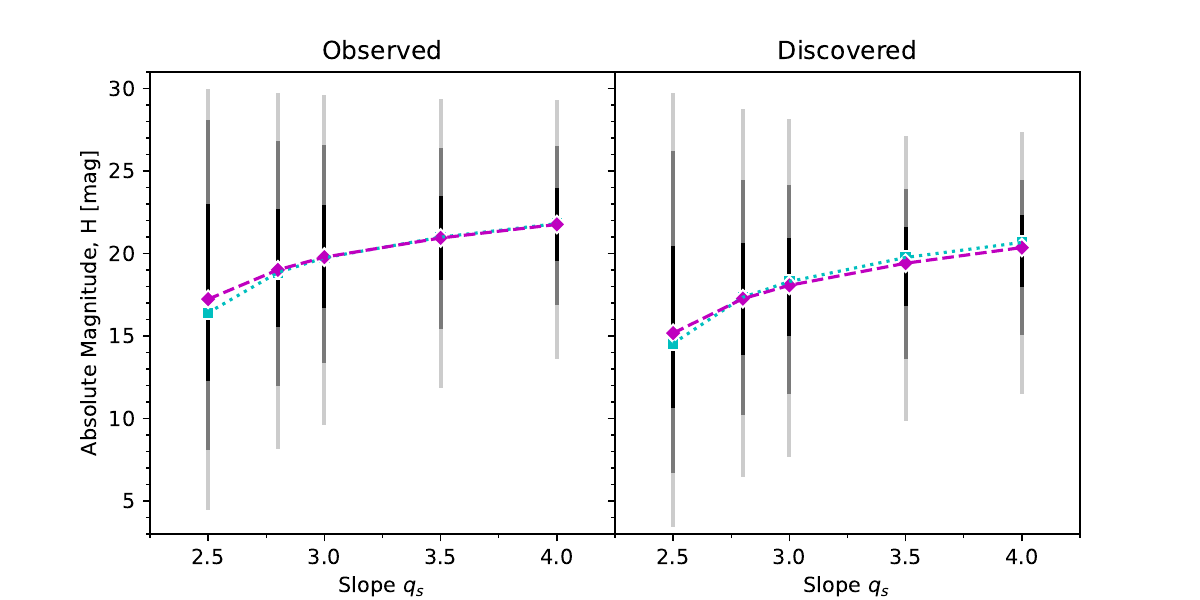}
    \caption{Summary statistics for the modelled absolute magnitude distributions of the observed (left) and discovered ISOs (right), for all considered values of slope $q_s$.
    A simple power law function with a sigmoid rollover was fitted to each binned distribution.
    The resulting $1\sigma$, $2\sigma$, and $3\sigma$ limits for each slope $q_s$ are indicated by vertical black lines of decreasing opacity.
    The mode (cyan squares) and median (magenta diamonds) values are also indicated.
    }
    \label{fig:h_dists}
\end{figure}

Summary statistics for the absolute magnitude probability density functions of observed and discovered ISOs in LSST are shown in Figure~\ref{fig:h_dists}.
As slope $q_s$ increases, the distributions become narrower and the median shifts to higher values of $H_r$, for both observed and discovered objects.
The most probable discoveries have $H_r=14.6\text{--}20.7$ for the slopes $2.5 \leq q_s \leq 4.0$. 
This result implies that the most frequently discovered absolute magnitude will be a tracer of the intrinsic distribution's slope parameter $q_s$.
Perhaps surprisingly, if \oneI{} was discovered in LSST, its absolute magnitude \citep[$H_r \sim 22.4$;][]{`OumuamuaISSITeam_2019} would be consistent with all of our model $H_r$ distributions within $2\sigma$.
On the contrary, discovering a single large ISO with $H_r \leq 10$ would strongly implicate a shallower slope ($q_s\leq3.0$) for the intrinsic absolute magnitude distribution.

From our simulated discovery rates, we predict that LSST will discover 6--51 ISOs (with the highest yield for $q_s=2.5$), for an assumed number density $n\left(H_r\leq22\right) \sim 0.1$~au$^{-3}$ \citep{Meech_2017, Do_2018}; see Appendix~\ref{sec:appendixExpectedNumbers} for the full calculation.
Caution is necessary when interpreting the number of successful discoveries in LSST.
The ISO discovery rate is linearly correlated with the spatial number density $n$, which has a large uncertainty \citep{Flekkoy_2023}.
Calibration of surveys for ISO detectability is challenging; it is possible that ISOs were present but not discovered.
The discovery rate is also non-linearly correlated with the intrinsic absolute magnitude distribution.
The result is a degeneracy, where a shallow absolute magnitude distribution slope and a high number density will produce the same number of LSST discovered ISOs as a steep absolute magnitude distribution slope and a low number density.
Thus the number of ISOs that LSST discovers will not independently constrain the luminosity function.
However, the distribution of $H_r$ values within the discovered sample may provide more information, as shown in Figure~\ref{fig:h_dists}.

\begin{figure}
    \centering
    \hfill
    \begin{minipage}[b]{.35\linewidth}
        \centering
        \begin{tabular}{c|ccccc}
            $q_s$ & 2.5 & 2.8 & 3.0 & 3.5 & 4.0 \\ \hline
            2.5 & \cellcolor{black!100}{} & \cellcolor{lightgray!100}{45} & \cellcolor{lightgray!50}{21} & 6 & 4 \\
            2.8 & \cellcolor{lightgray!50}{45} & \cellcolor{black!100}{} & \cellcolor{lightgray!100}{174} & \cellcolor{lightgray!50}{15} & 8 \\
            3.0 & \cellcolor{lightgray!50}{22} & \cellcolor{lightgray!100}{173} & \cellcolor{black!100}{} & \cellcolor{lightgray!50}{26} & \cellcolor{lightgray!50}{11} \\
            3.5 & 7 & \cellcolor{lightgray!50}{16} & \cellcolor{lightgray!50}{27} & \cellcolor{black!100}{} & \cellcolor{lightgray!50}{74} \\
            4.0 & 4 & 8 & \cellcolor{lightgray!50}{11} & \cellcolor{lightgray!50}{73} & \cellcolor{black!100}{} \\
        \end{tabular}
        \captionof{table}{The number of discovered ISOs required to reject the hypothesis that their $H_r$ values are drawn from the same absolute magnitude distribution as a given `model' distribution.}
        \label{tab:AD_test_H}
        \vspace{7em}
    \end{minipage}
    \hfill
    \begin{minipage}[b]{.625\linewidth}
        \centering
        \includegraphics[width=\linewidth]{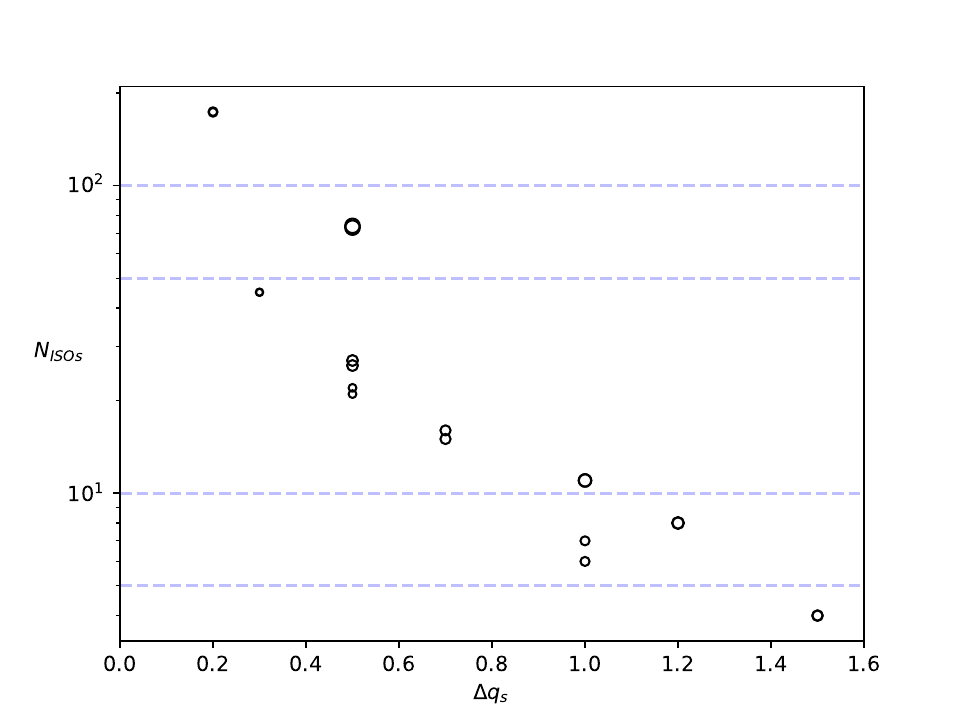}
    \end{minipage}
    \captionof{figure}{Comparison of the difference in slope $q_s$ of the two compared model $H_r$ distributions to the number of discoveries required to reject models.
    The size of each data point is proportional to the product of the compared slope values.
    The blue horizontal dashed lines indicate sample sizes of 5, 10, 50 and 100 ISOs.
    Fewer discovered ISOs are needed to distinguish between slope values that are further apart.
    For the same $\Delta q_s$, steeper slopes require more discoveries in order to be distinguished.}
    \label{fig:AD_H}
\end{figure}

We investigated whether the LSST sample of ISOs could constrain the intrinsic absolute magnitude distribution by performing a power analysis.
Across $10^3$ trials, we randomly drew two sets of $H_r$ values of size $N$ from different discovered ISO absolute magnitude distributions, and compared them to both models using an Anderson-Darling test \citep{Anderson_1952}.
The minimum sample size $N$ required for the incorrect model to be rejected for 95\% of the trials (representing a 95\% confidence interval) is presented in Table~\ref{tab:AD_test_H} and visually illustrated by Figure~\ref{fig:AD_H}.
A negative correlation exists between the number of discoveries needed and the difference in the slope values being compared.
For the same slope difference $\Delta q_s$, more ISOs are needed to differentiate between two steeper slopes (e.g. $q_s=3.5$ and $q_s=4.0$) than two shallower slopes (e.g. $q_s=2.5$ and $q_s=3.0$).
The smallest expected sample size from LSST is for the steepest slope, $q_s=4.0$.
Given six discoveries, the absolute magnitude distribution slope can be constrained to within $\simeq \pm 1.5$ of the intrinsic value.
In comparison, for the largest expected sample size of 51 ISOs for the shallowest slope $q_s=2.5$, the absolute magnitude distribution slope can be constrained to within $\simeq \pm 0.3$ of the intrinsic value.
These outcomes suggest that the predicted LSST sample of ISOs will strongly aid the characterisation of the Galactic population, regardless of the sample size.

\begin{figure}
    \centering
    \includegraphics[width=\linewidth]{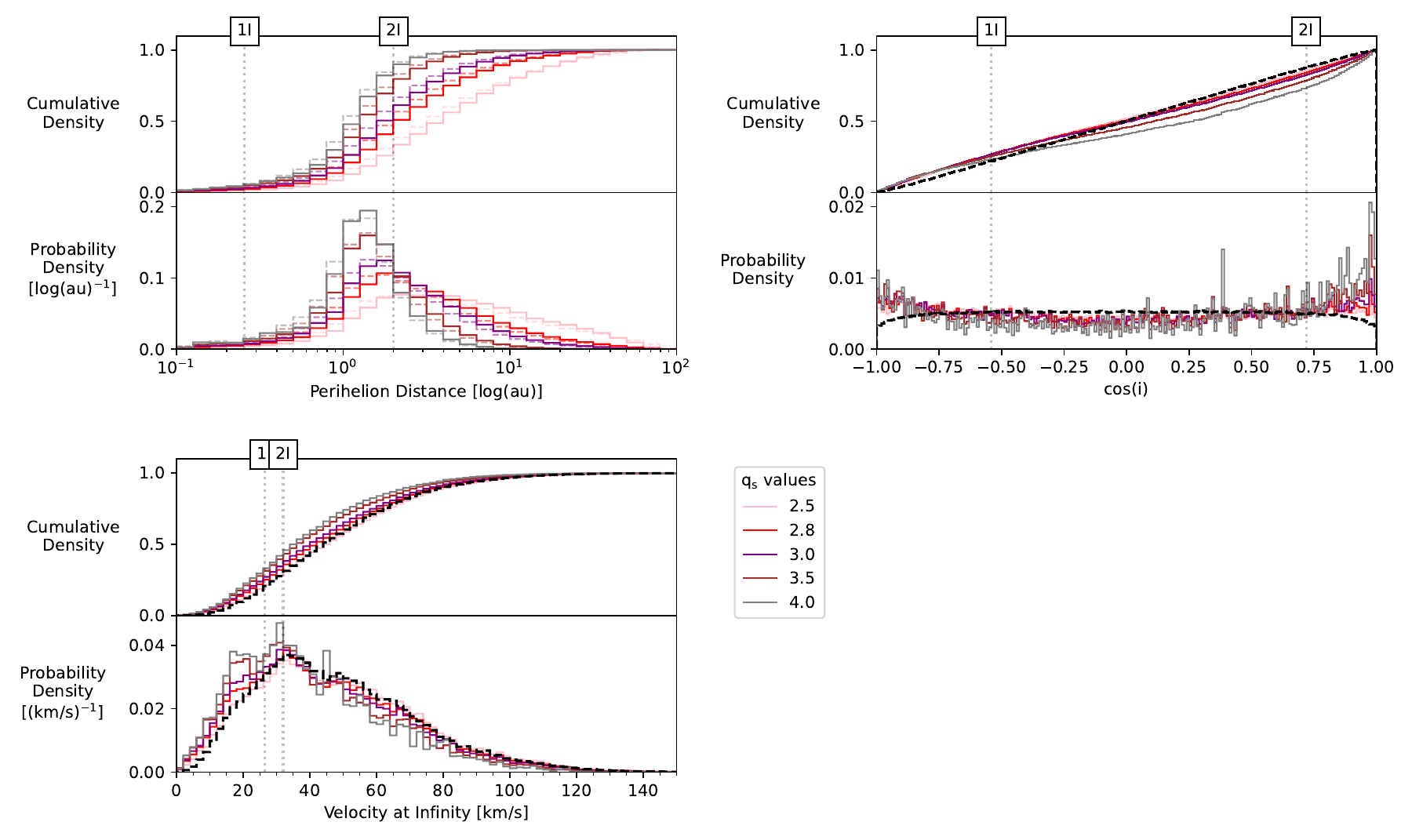}
    \caption{
    Cumulative and probability density functions for the perihelion (top left), cosine of inclination (top right) and velocity at infinity (bottom) for ISOs discovered in LSST.
    Comparisons are made to the observed ISO sample (coloured dashed curves) for perihelion and the \OO{} model (black dashed curve) for inclination and velocity.
    Vertical dotted black lines indicate the corresponding values for the two known ISOs.}
    \label{fig:qiv}
\end{figure}

Other characteristics of the ISOs discovered by LSST, such as perihelia, inclination, and velocity, will also help constrain the intrinsic ISO absolute magnitude distribution slope.
Figure~\ref{fig:qiv} demonstrates that the perihelia and inclination distributions of ISOs in LSST will depend on the intrinsic absolute magnitude distribution. 
For a shallower intrinsic slope $q_s=2.5$, LSST's ISOs will have modal perihelion $q\sim4$~au and an equal proportion of prograde and retrograde orbits.
In contrast, for the steepest slope $q_s=4.0$, the modal perihelion will be $q\sim1$~au and the LSST sample will have disproportionally prograde orbits.
Although LSST will have a deeper $5\sigma$ limiting magnitude than the Pan-STARRS survey in which \oneI{} was detected \citep[$m_r\sim23$ compared to $\sim21$;][]{Denneau_2013}, we find that similar-sized objects to \oneI{} ($H_r\sim22.4$) are unlikely to contribute to the LSST sample, with probability $P\sim0.01$ --- we can expect at most one \oneI{}-like `asteroidal' ISO in LSST, for current spatial density estimates.
This is an order of magnitude less probable than inactive objects of dimensions comparable to the nuclear diameter of \twoI{}.
There is also a discernable difference in the expected median velocity at infinity for the different absolute magnitude distribution slopes.
The steepest slope $q_s=4.0$ results in slower ISOs discovered in LSST, with a median of $v_\infty \sim 35$~km/s compared to $v_\infty \sim 45$~km/s for the shallowest slope $q_s=2.5$.
Interestingly, both known ISOs had $v_\infty$ consistent with steeper slopes.

\begin{figure}
    \centering
    \includegraphics[width=\linewidth]{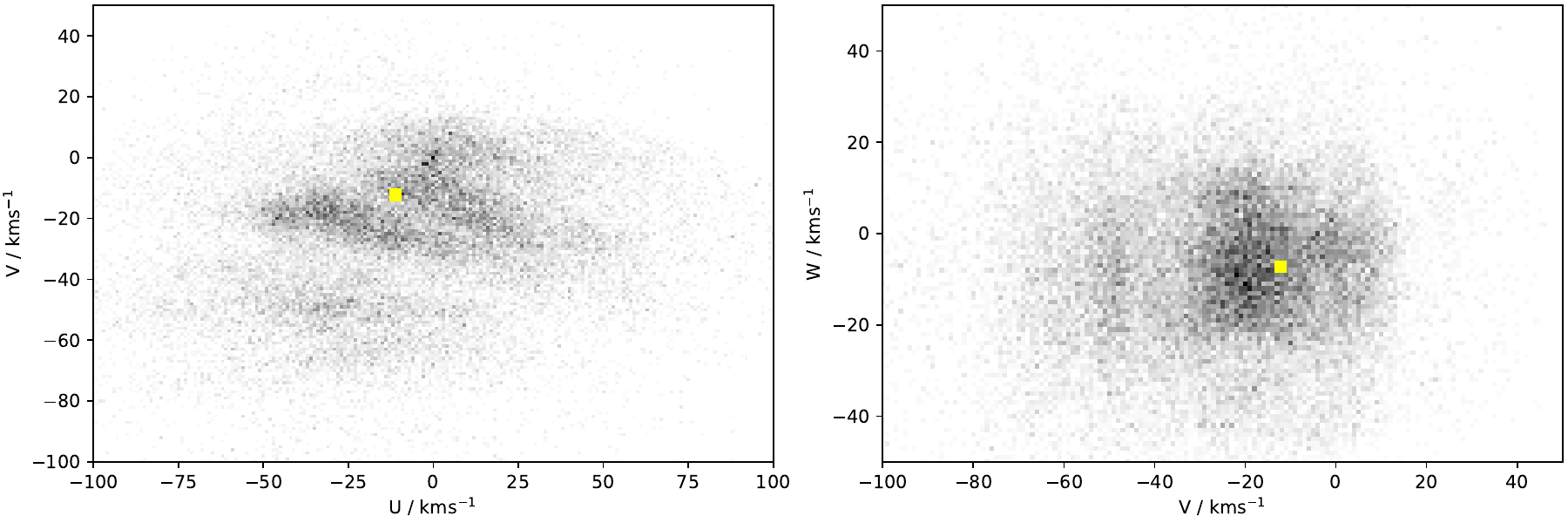}
    \caption{Probability density functions for the Galactic velocity distributions of the LSST discovered ISOs from the $q_s=2.8$ intrinsic absolute magnitude distribution.
    The yellow square denotes the Local Standard of Rest.
    The structures predicted by \citet{Hopkins_2025} are still apparent in the velocity distribution the discovered ISOs will be drawn from, including the over-density at the Sun's velocity $\left(U,V,W\right)=\left(0,0,0\right)$ due to gravitational focusing.}
    \label{fig:UVW}
\end{figure}

The \OO{} model allows us to explore several properties of the LSST ISO sample: the age of the ISOs (which in the model is held contemporaneous with the star's age, assuming an early unbinding event), the metallicity of their origin stars, and the water-mass fraction of the ISOs assuming all formed beyond the ice line, which correlates with stellar metallicity.
We see no relationship between these properties and the $H_r$-distribution; the LSST sample is effectively unbiased on these parameters. 
The ISO pre-encounter velocity structures predicted in \citet{Hopkins_2025} from Galactic dynamical structures are also recovered for the LSST sample, in both $\left(U,V\right)$ and $\left(V,W\right)$ phase space, although slightly less pronounced (Fig.~\ref{fig:UVW}).
The over-density at the Sun's velocity $\left(U,V,W\right)=\left(0,0,0\right)$ caused by gravitational focusing is also discernable.

\section{Structural patterns in ISO discoveries}
\label{sec:biases}

\begin{figure}[]
    \centering
    \includegraphics[width=\linewidth]{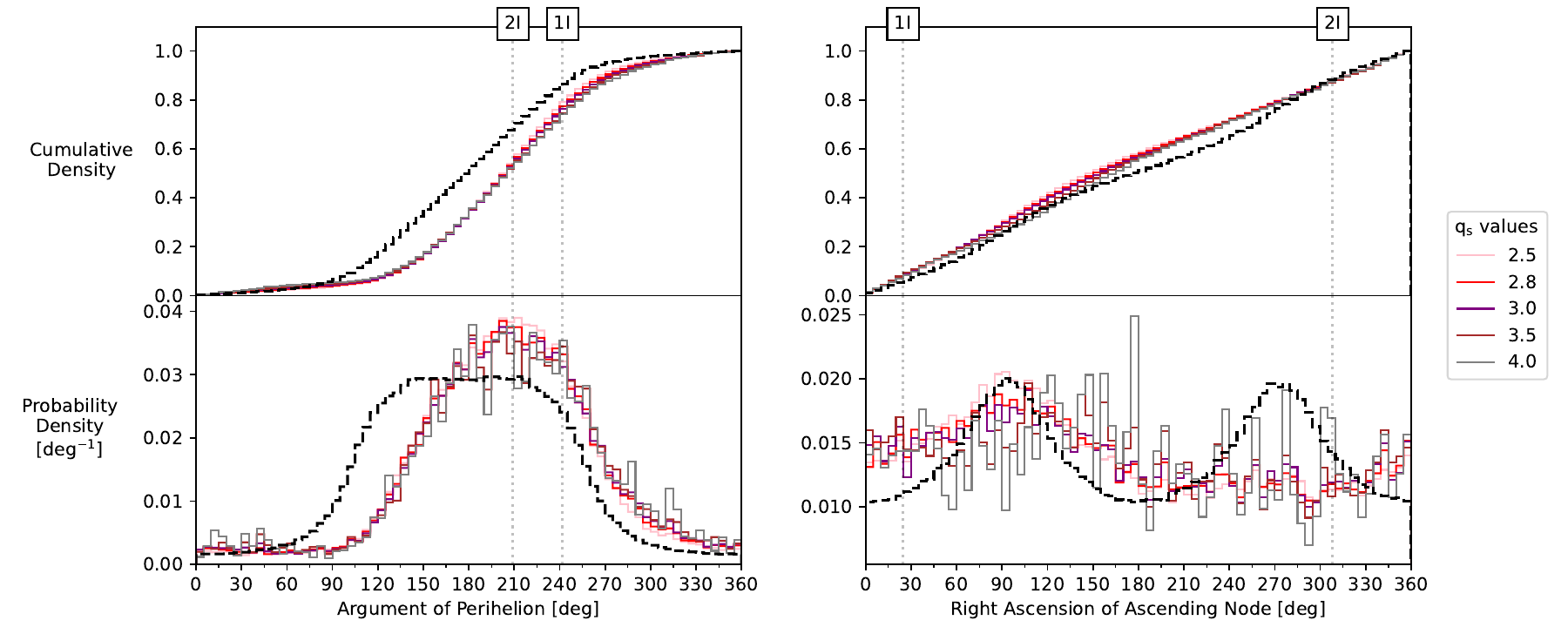}
    \caption{Same as Fig.~\ref{fig:qiv}, but for argument of perihelion (left) and right ascension of the ascending node (right).
    The dashed black curves indicate the cumulative and probability density functions for the intrinsic model (the \OO{} population).
    }
    \label{fig:om_Om}
\end{figure}

The survey cadence and discovery choices will uniquely shape the properties of the LSST ISO sample.
The orbital orientations of the LSST ISO sample will be biased due to the telescope's location and the survey footprint.
The Simonyi Survey telescope is located on Cerro Pach\'{o}n (Chile) in the Southern Hemisphere; as a result, the survey footprint is restricted to fields with declinations $\delta \lesssim 35^\circ$.
This means that most ISOs with perihelia in the Northern Hemisphere will be outside of the survey footprint when they are at their brightest.
The argument of perihelion and right ascension of the ascending node distributions for the LSST sample demonstrate the significance of this Southern Hemisphere survey bias (Fig.~\ref{fig:om_Om}).
Orbits with $0<\omega<90^\circ$ or $\Omega \sim 270^\circ$ are distinctly under-represented in the LSST sample compared to the \OO{} model.
Interestingly, \oneI{} and \twoI{} have arguments of perihelia near the peak probable value of LSST discoveries.

\begin{figure}[]
    \centering
    \includegraphics[width=\linewidth]{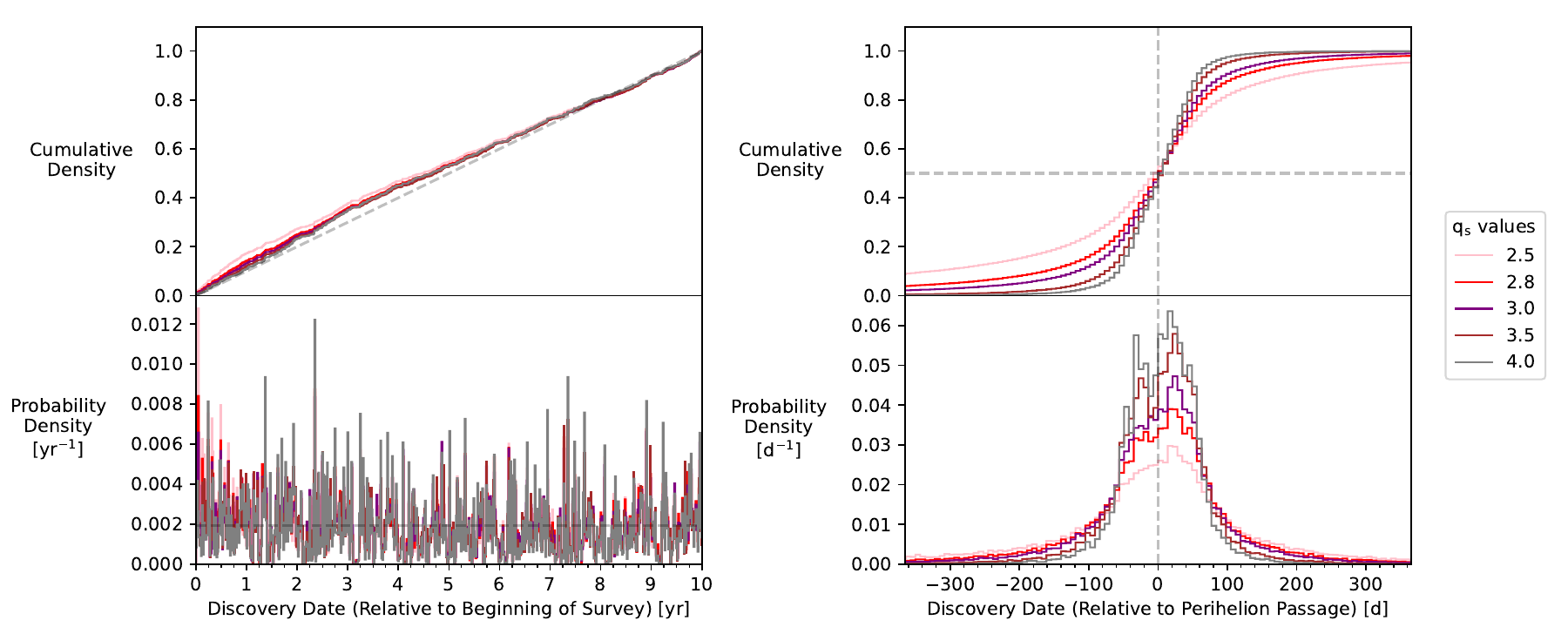}
    \caption{Same as Fig.~\ref{fig:qiv}, but for discovery date relative to the start of the survey (left) and to the time of perihelion passage (right).
    The dashed lines indicate a uniform discovery rate (left), the time of perihelion passage (right; vertical line) and the 50\% cumulative density (right; horizontal line).}
    \label{fig:disc_rel}
\end{figure}

The discovery circumstances of ISOs in LSST will be contrary to expectations for other Solar System small-body populations.
Previous modelling efforts have shown that most LSST discoveries of bound Solar System objects are made in the first few years of the survey \citep{LSST_R1}.
A similar prediction could be naively applied to ISOs --- when LSST reaches first light it might first discover the ISOs in the initial observable survey volume, until an equilibrium discovery rate is reached when the refreshing of the local ISO population from more distant points becomes the dominant driver of the discovery rate.
However, our simulations demonstrate that ISOs will be discovered at an almost constant rate throughout the survey (Fig.~\ref{fig:disc_rel}).
We also investigated whether there was a correlation between seasonal effects (e.g. solar conjunction, annual weather patterns etc) and the discovery rate of ISOs.
We found that inter-yearly ISO discoveries are also uniformly distributed in time, accounting for leap years and regularly scheduled maintenance downtime of variable length.
Thus we interpret the linear discovery of ISOs in LSST as a consequence of the assumed uniform refresh rate of ISOs entering and leaving their observable sphere, rather than specific cadence choices or annual Sun-Earth geometries. 
Additionally, ISOs will be discovered before and after perihelion with equal probability (Fig.~\ref{fig:disc_rel}); however, the most probable scenario is discovery $\sim20$~days after perihelion for all slopes $q_s$.
This contradicts previous predictions that ISOs will be predominantly discovered pre-perihelion \citep{Hoover_2022}.

Contrary to expectation \citep[e.g.][]{Schwamb_2023}, the twilight microsurvey does not provide critical discovery of asteroidal ISOs.
Our simulations show that ISOs discovered only in the twilight survey typically have absolute magnitudes $17 \leq H_r \leq 20$ with preferentially lower perihelia and velocities at infinity than objects discovered in the main survey.
In all other characteristics considered in this work (both orbital and physical), twilight discovered ISOs are consistent with objects found in the main survey.
In reality, the twilight ISO discoveries make up 0.9--1.4\% of all discoveries for any slope $q_s$, and a further 0.3--2.2\% of discovered ISOs are found by both the twilight and main surveys.
Thus, for our predicted number of LSST ISOs, we can expect at most one ISO discovery from the twilight survey (with $\sim70\%$ probability) and about one object discovered in both surveys, which occurs for the shallowest absolute magnitude slope $q_s=2.5$.
For all twilight discoveries, $\sim54\text{--}65\%$ are discovered in evening twilight, with the slope $q_s=2.5$ incurring the highest proportion of morning twilight discoveries.

\begin{figure}
    \centering
    \includegraphics[width=\linewidth]{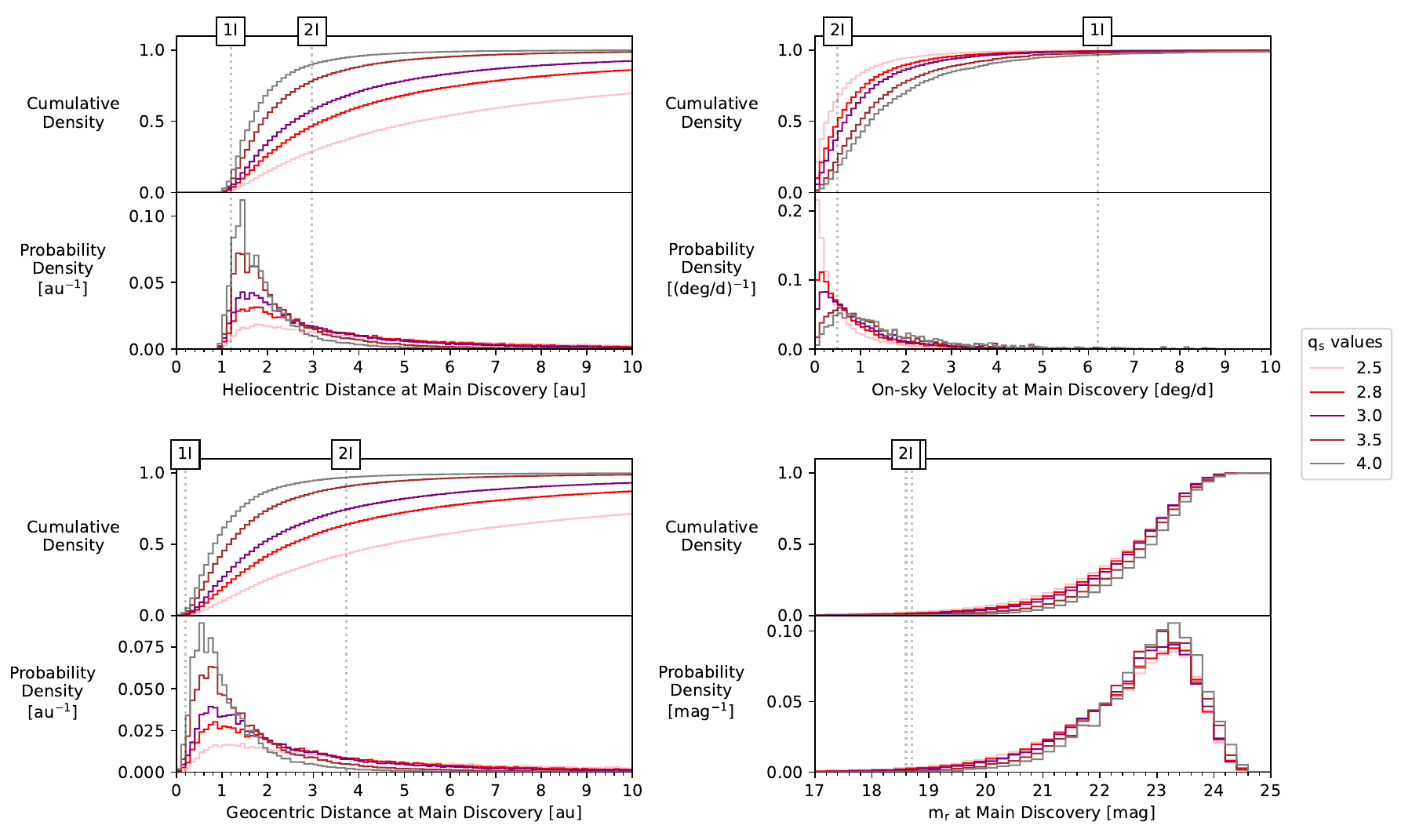}
    \caption{Same as Fig.~\ref{fig:qiv}, but for the properties of ISOs at the time of their discovery in the main survey of LSST. Top left to bottom right: heliocentric distance, on-sky velocity, geocentric distance, and apparent $r$-band magnitude $m_r$.
    Vertical dotted black lines indicate the corresponding values for the two known ISOs at the time of their discovery; note that \oneI{} was discovered by NEO discovery survey Pan-STARRS, while \twoI{} was discovered at low solar elongation and was cometary.}
    \label{fig:atDiscovery}
\end{figure}

Our simulations suggest that ISOs will be discovered at small distances and velocities.
ISOs found in the main survey typically have heliocentric distances $\rh \sim 1\text{--}3$~au, geocentric distances $\Delta \sim 0\text{--}3$~au and on-sky rates of motion $\mu \sim 0\text{--}2$~$^\circ$~d$^{-1}$ at discovery (Fig.~\ref{fig:atDiscovery}).
This is broadly consistent with previous results from \citet{Cook_2016} and \citet{Hoover_2022}, which predicted that ISOs would be discovered with perihelia $q\approx1$~au (contrast Fig.~\ref{fig:qiv}) and pass within $\sim1.5$~au of the Earth.
The distributions of distance and velocity at discovery for our simulated \OO{} ISOs also place into context the nature of the first ISO; \oneI{} was discovered comparably close to Earth ($\Delta \sim 0.2$~au, but in a shallower survey), though with an atypically large on-sky rate of motion ($\mu \sim 6$~$^\circ$~d$^{-1}$).
While most ISOs discovered by LSST will be found near the Earth, only a minority will be discovered with motion rates as extreme as \oneI{}.
The distributions of distance and velocity at discovery are also dependent on the intrinsic absolute magnitude distribution.
Objects from a steeper slope distribution are typically discovered at lower heliocentric and geocentric distances and higher on-sky velocities than from a shallow slope distribution.

The most probable $r$-band apparent magnitude of ISOs at discovery is $\sim23.3$~mag.
Above this value, the likelihood of discovery decreases rapidly with $m_r$; no ISOs are discovered with apparent magnitude $m_r \leq 25$.
This limiting magnitude for ISO discovery in LSST is ${\sim4.7}$ magnitudes fainter than the apparent magnitudes of the two known ISOs at their respective discoveries, as expected given the larger-aperture facility.
In conjunction with the distances of ISOs at discovery, this confirms that LSST will find ISOs within a larger detectable volume than previous surveys.

\begin{figure}[]
    \centering
    \includegraphics[width=\linewidth]{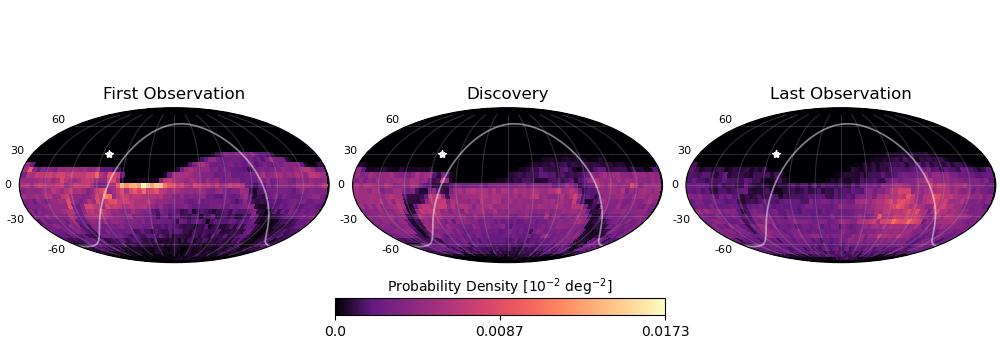}
    \caption{2-dimensional illustrations for the on-sky locations ($\alpha$, $\delta$) for the first, discovery, and last images of the LSST discoveries for $q_s=2.8$ for all years of the survey.
    Other absolute magnitude slopes also show similar trends.
    The white star indicates the Solar Apex and the thin white line indicates the Galactic plane.}
    \label{fig:onsky_FDL_allH_q28}
\end{figure}

In addition to the observation scheduling within the LSST footprint, the intrinsic dynamics of the galactic ISO population affect the on-sky location of an ISO observed by LSST.
At first observation, ISOs tend to be found in the northern area of the LSST footprint, nearest to the Solar Apex (Fig.~\ref{fig:onsky_FDL_allH_q28}).
The Solar Apex is the direction of motion of the Sun relative to the Local Standard of Rest \citep{Schonrich_2010}, located at approximately $(\alpha,\delta)=(\mathrm{18h}, +30^\circ)$ in the Northern celestial hemisphere \citep{Ridpath_2012}.
This is the direction from which ISOs are expected to encounter the Solar System \citep{McGlynn_1989, Stern_1990}.
Compared to their first observation, ISOs are then discovered almost uniformly across the LSST footprint.
The majority of discovered ISOs (95\%) have arc length in LSST $\geq4-17$~days (median of $35-200$~days); the steepest absolute magnitude slope $q=4.0$ has the shortest visibility. 
Their last observation is most likely nearest the Solar Antapex (Fig.~\ref{fig:onsky_FDL_allH_q28}), typically $\geq10$~days after their first observation in LSST for 95\% of ISOs from the $q_s=2.8$ distribution and $\geq100$~ days for 50\% of discoveries.

\begin{figure}[]
    \centering
    \includegraphics[width=0.5\linewidth]{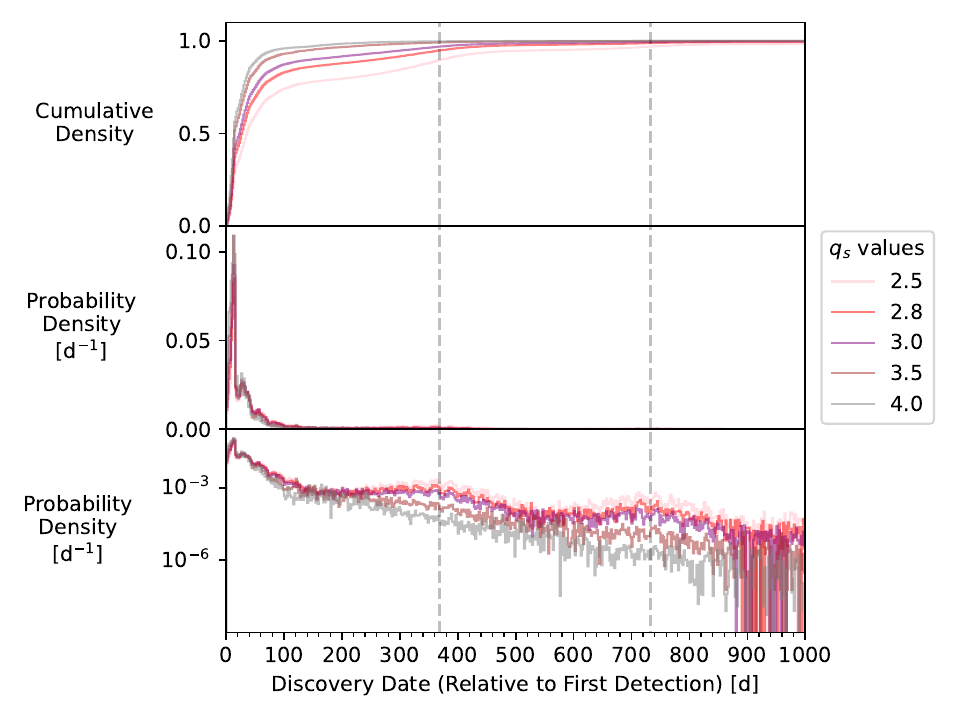}
    \caption{Same as Fig.~\ref{fig:qiv}, but for discovery date relative to first detection. The dashed grey vertical lines indicate 380 and 745 days respectively, corresponding to integer years plus 15 days.}
     \label{fig:cp_discovery_rel_detection}
\end{figure}

\begin{figure}[]
    \centering
    \includegraphics[width=\linewidth]{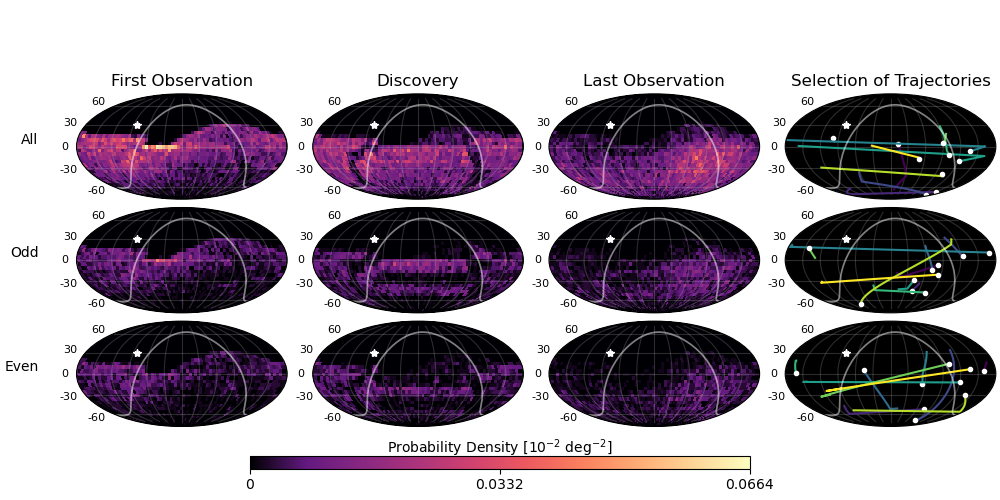}
    \caption{2-dimensional illustrations for the on-sky locations ($\alpha$, $\delta$) for the first, discovery, and last images of the LSST discoveries with $13 \leq H \leq 14$ for $q_s=2.8$ for all years (top row), odd rolling years (middle row) and even rolling years (bottom row).
    This excludes Year 2 of the simulation, which only deploys a rolling cadence over half of the survey footprint.
    The white star indicates the Solar Apex and the thin white line indicates the galactic plane.
    A sample of the observations for 10 ISOs is shown in the last column: the first, discovery, and last observations are connected by straight lines of random colour and the last observation is marked by a white dot to indicate the direction of motion.}
    \label{fig:onsky_FDL_H13_q28}
\end{figure}

The rolling cadence strategy within the survey causes distinct patterns in the LSST ISO discoveries.
While most ISOs are discovered within $\sim100$~days of their first observation in LSST, some ISOs are found over a year later (Fig.~\ref{fig:cp_discovery_rel_detection}).
The probability of discovery at ${>}100$~days after initial detection also appears to oscillate in log space over one year, reaching maxima at ${\sim365n+15}$ days after the first detection (where $n$ is an integer number of years).
This is a result of the rolling cadence strategy; an object may be observed in an ``off" stripe one year but is insufficiently sampled to be discovered and is subsequently discovered by LSST the following year once the stripe rolls ``on".
Figure~\ref{fig:onsky_FDL_H13_q28} demonstrates this within the context of the LSST footprint.
ISOs discovered in rolling years are first observed across several areas on the celestial sphere; i.e. they are not necessarily constrained to being observed and discovered in one stripe only.
Additionally, ISOs are slightly more discoverable in odd rolling years than even rolling years.
This is due to odd years surveying the more northern rolling stripe pattern.
Examining a selection of ISO on-sky trajectories from their first, discovery and last observations shows that the general on-sky motion of ISOs is from north to south, and from the galactic plane inward of the Solar circle to the galactic plane outward of the Solar circle.

\section{Characterising LSST's ISOs with follow-up}\label{sec:CWs}

After their discovery in LSST, ISOs will be a high priority for follow-up characterisation.
While LSST will provide some incidental self-follow-up, there is limited capacity for ISO-targeted observations by the Simonyi Survey Telescope \citep{Andreoni_2024}.
Instead, the responsibility of ISO follow-up will fall on telescope facilities around the world. 
Observational campaigns requesting telescope resources typically take hours to days (for Director's Discretionary time) to months (for regular semester time) to be approved, so it is critical to know whether the target ISO will have reduced in brightness by the time the observations are approved and scheduled.

We investigate the length of time after discovery that an LSST-discovered ISO will be bright enough to be targeted by various facilities.
We use two indicative magnitude limits: $m_r=23$ for colours or spectroscopic instruments \citep[e.g. Gemini Multi-Object Spectrograph at Gemini Observatory;][]{Dotto_2003}\footnote{For instance, \oneI{} was predicted to have $m_r = 22.7$ during the observations in \citet{Bannister_2017}.}
and $m_r=28$~mag for space telescopes \citep[e.g. HST;][]{HST_WFC3_Handbook}.
Since an ISO will generally have an initial multi-day arc from its discovery images (single-day arcs are limited to the small fraction of twilight survey discoveries), we assume that the object's on-sky velocity would be known throughout the rest of its passage through the Solar System to a level of accuracy suitable for non-sidereal observations with small-FOV instruments.
Therefore, we ignore trailing losses in the following analysis.
For the rest of this section, an ISO's `apparent magnitude' refers to its untrailed apparent magnitude within a non-sidereally-tracked follow-up context, unless stated otherwise.

Due to the time-bounded nature of surveys, when LSST begins surveying, some ISOs will already be outbound from the Solar System while others are just on their way into their observable sphere.
This means that the ISO discoveries in the earliest survey years of LSST are a combination of objects eventually reaching an observable magnitude and objects that would have been discovered already if the survey had started earlier.
This phenomenon is both orbit- and luminosity-function dependent, as the amount of time an object spends at a magnitude observable by LSST is related to its perihelion distance and absolute magnitude.
The largest objects ($H_r=5$) can have residence times on the order of decades, thus contributing to this effect.
However, these are also far less frequent in the intrinsic and discovered populations (regardless of the absolute magnitude distribution chosen in this work) so the impact should be minor.
Nevertheless, we analyse only ISO discoveries from Year 2 to Year 6 within the simulated 10-year survey to mitigate this as much as possible.
For each discovered ISO, we evaluate its apparent magnitude until it becomes fainter than $m_r=28$~mag.
Additionally, we only consider objects from the $q_s=2.8$ absolute magnitude distribution for the following analysis.

\begin{figure}[]
    \centering
    \includegraphics[width=\linewidth]{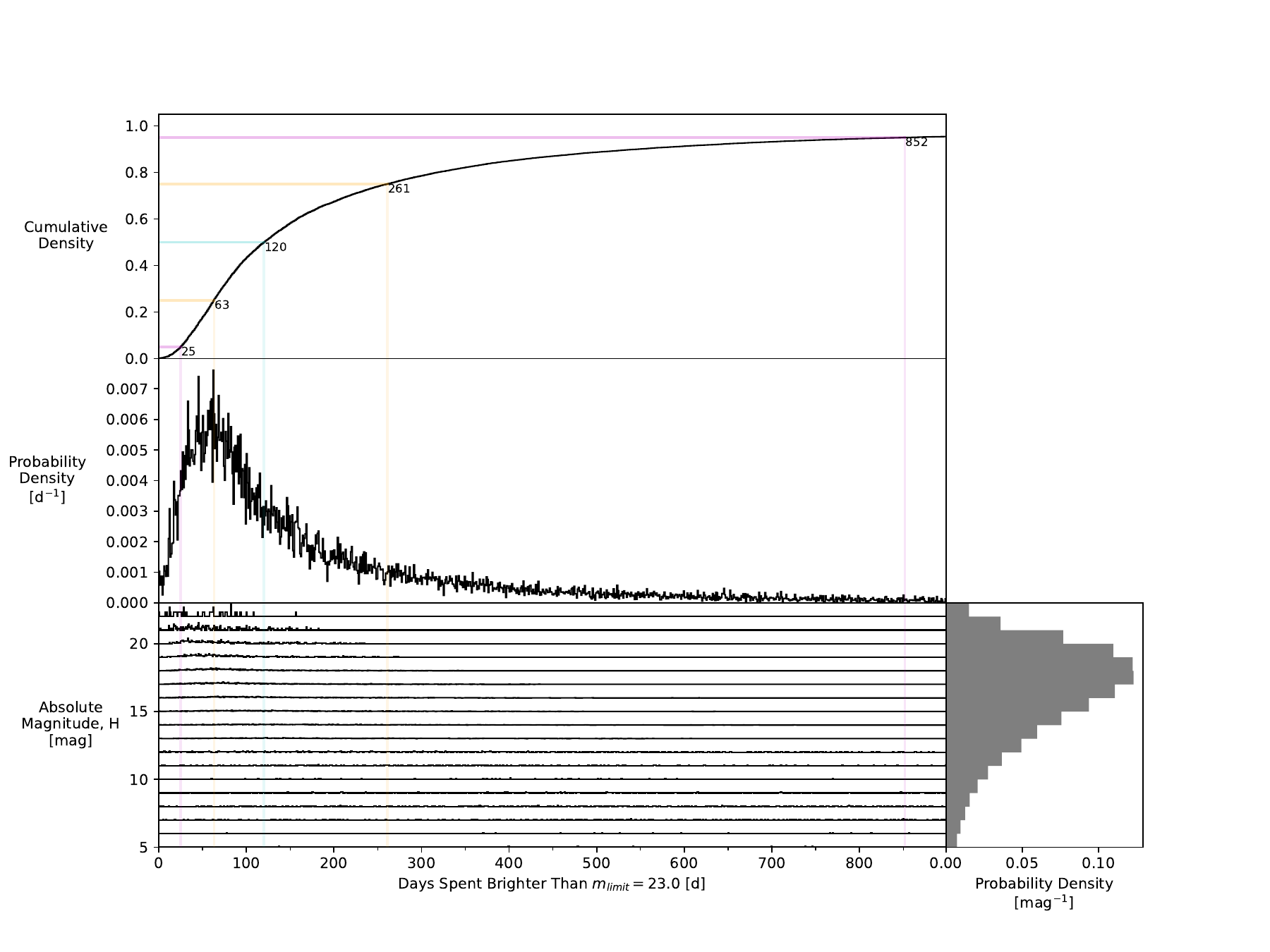}
    \caption{Cumulative and probability density functions for the amount of time an ISO will spend during its Solar System passage above the magnitude limit for spectroscopic observation if the object is discovered brighter than $m_r=23$ ($q_s=2.8)$.
    Different coloured lines indicate percentiles of interest: 5\textsuperscript{th} and 95\textsuperscript{th} (magenta), 25\textsuperscript{th} and 75\textsuperscript{th} (orange), and 50\textsuperscript{th} (cyan).
    Bottom left: qualitative probability density functions for each magnitude interval. This plot type we term a ``tally plot".
    Bottom right: probability density function for the $H_r$ distribution of discovered ISOs.
    Middle: probability density function for the whole population, produced by weighting each probability density function in the tally plot (bottom left) by the corresponding $H_r$ probability density value (bottom right).
    Top: cumulative probability density function for the probability density function in the middle panel.}
    \label{fig:cpt_q28_DBTM23_already}
\end{figure}

\begin{figure}[]
    \centering
    \includegraphics[width=\linewidth]{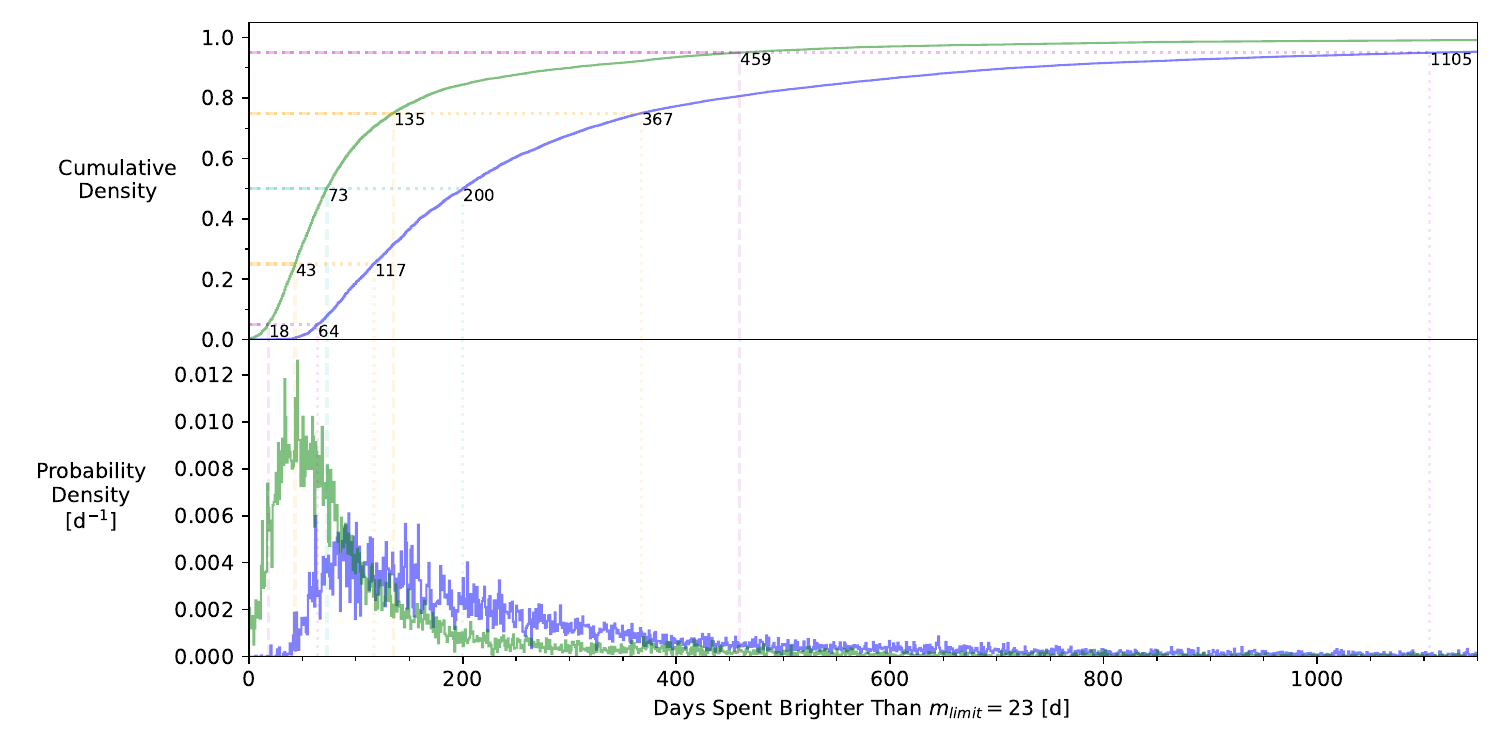}
    \caption{Cumulative (top) and probability density functions (bottom) for ISOs from the $q_s=2.8$ absolute magnitude distribution discovered inbound (blue) or outbound (green).}
    \label{fig:cpt_q28_DBTM23_already_ivo}
\end{figure}

Most ISOs are discovered already bright enough for spectroscopic observations.
This is $H_r$-dependent; as $H_r$ increases, the proportion of objects that never or eventually brighten to $m_r=23$ after their discovery decreases roughly linearly.
Objects are discovered above and below the $m_r=23$ limit due to differences in their orbital position relative to perihelion passage.
As expected, most objects fainter than $m_r=23$ at discovery are on inbound trajectories, while those that are never brighter than $m_r=23$ after discovery are on outbound trajectories.
Objects brighter than $m_r=23$ at discovery are equally likely to be on the inbound or outbound portion of their orbit.
Smaller ISOs ($H_r\geq17$) are almost always discovered above the $m_r=23$ threshold.
As a result, 93.4\% of the discovered ISOs will be initially brighter than $m_r=23$, while 4.0\% and 2.7\% never or eventually reach this magnitude after their discovery, respectively.
Of the ISOs bright enough for spectroscopic observations immediately, 95\% will be observable for $\geq25$~days (Fig.~\ref{fig:cpt_q28_DBTM23_already}); this is also true for the most probable discovered ISO with $17 \leq H \leq 18$.
Inbound objects are observable $\sim2\text{--}3$ times longer than outbound objects; the longest observability period for the 95\% percentile is $\geq 1105$~days compared to $\geq 459$~days (Fig.~\ref{fig:cpt_q28_DBTM23_already_ivo}).

All ISOs are brighter than the magnitude limit $m_r=28$ for space-based telescope follow-up for at least 30 days after their discovery.
Most (95\%) will be viable for follow-up for $\geq160$~days; this increases to $\geq199$~days for the most probable discovery ($17 \leq H \leq 18$).
In rare cases (95th percentile), ISOs are observable for $\geq3084$~days, or $\sim8.5$~years.
Inbound objects are observable $\sim1.2\text{--}1.6$ times longer than outbound objects; the longest observability period for the 95\% percentile is $\geq 3503$~days compared to $\geq 2434$~days ($\sim 9.5$~years and $\sim 6.5$~years respectively).
    
\section{Discussion}
\label{sec:discussion}

ISOs are a cosmogonically unique and observationally disadvantaged small-body population.
While they may hold the answers to many questions about the fundamental physics of planet formation, they are rarely discovered.
In this survey simulation of ISOs in the upcoming LSST, we quantify the observational and discovery biases for the \OO{} population model, under realistic assumptions about ISO physical properties.
From our simulated discovery rate of ISOs in LSST, we estimate that 6--51 ISOs will be discovered during LSST.
This is in agreement with works such as \citet{Cook_2016} and \citet{Trilling_2017}, and is lower than other studies which omit trailing losses, LSST cadence details and/or LSST SSP discovery criteria \citep{Rice_2019, Hoover_2022, Marceta_2023a}.
The number of discoveries we expect is also sufficient to test the prediction of \citet{Hopkins_2025} that the \OO{} model we use here will be distinguishable (N=20 for $2\sigma$ or N=50 for $3\sigma$ difference) from OB-type smooth Gaussians (cf. Fig.~\ref{fig:UVW}).

\subsection{The Unobservability of ISOs}
\label{sec:unobservability}

ISOs are extremely difficult to observe and discover in Solar System surveys such as LSST due to a unique array of observational biases.
A fundamental and invariable characteristic of ISOs is their single perihelion passage through the Solar System.
Unlike the majority of Solar System small body populations, ISO discovery does not have the luxury of repeated attempts; like long-period comets, their single passage is the only opportunity to detect them.

In addition to their single passage, an ISO's path within its observable sphere plays an important role in its discovery.
The size of an ISO's heliocentric observable sphere depends on its absolute magnitude and the limiting magnitude of LSST.
For an ISO to be observable by LSST, the object must have a perihelion distance within this sphere, i.e. $q\leq r_{\mathrm{sph}}$ (Eq.~\ref{eq:r_max_obsSphere}).
The most probable perihelion for an observable ISO is $q \approx r_{\mathrm{sph}}$ rather than $q \ll r_{\mathrm{sph}}$; therefore the amount of time an ISO spends observable and the number of observations it receives is likely small.
This is confirmed by the orbital parameters of the discovered ISOs ($\S$~\ref{sec:sfd}).
ISOs are most likely to be discovered in the main survey with $4-6$ tracklets (depending on the intrinsic absolute magnitude distribution), which is almost the minimum requirement for discovery ($\S$~\ref{sec:detectability_definition}).
Discovered ISOs tend to have lower velocities than the model population (Fig.~\ref{fig:qiv}), in agreement with \citet{Cook_2016}, and higher residence times compared to the intrinsic population, which allows them to remain observable long enough to be discoverable.
This introduces an inclination bias, as prograde orbits have longer residence times than retrograde orbits of a similar orientation and shape.
Our analysis agrees with the expectation of \citet{Cook_2016}; to first-order, the inclination of LSST's ISOs will be uniform.
However, our simulations predict a bias toward prograde orbits which increases with absolute magnitude slope (Fig.~\ref{fig:qiv}).
This is contrary to the finding of \citet{Marceta_2020} that LSST will be biased toward retrograde orbits, potentially due to our accounting for trailing losses, survey cadence, and moving object discovery criteria.

At least 50\% of ISOs will be discovered with perihelia $q\leq5$~au for absolute magnitude distributions with slope $q_s\geq2.5$ (Fig.~\ref{fig:qiv}).
This means a significant fraction of ISOs may also be discovered with perihelia beyond Jupiter's orbit.
\citet{Cook_2016} and \citet{Hoover_2022} predicted that ISOs will be discovered with perihelia $q\approx1$~au and pass within $\sim1.5$~au of the Earth.
Our simulations disagree; ISOs with perihelia $q>2.5$~au will make up $10-70\%$ of LSST discoveries, depending on the absolute magnitude distribution.
The median $H_r$ of our discoveries ranges from 15.2 to 20.4 mag for $2.5 \leq q_s \leq 4.0$ (Fig.~\ref{fig:h_dists}).
Assuming an example albedo of $p=0.05$, this corresponds to objects of diameter $D\sim $0.5--5.4~km.
This is larger than the mean ISO size predicted by \citet{Marceta_2023a}; however, their discovery criteria only required three observations for an object to be discovered.

Other factors that we do not model in this work may also impact the observability of ISOs in LSST.
One known observational bias for at least some of the ISO population is due to light curve variability.
\oneI{}, the only asteroidal ISO discovered so far, demonstrated extreme fluctuations in brightness due to its shape and rotation rate \citep{`OumuamuaISSITeam_2019} which impacted its observability \citep{Levine_2023}.
While the implementation of an ISO population shape model is left for future work, preliminary simulations of \oneI{}-like objects --- which we parameterise with absolute magnitude $22 \leq H_r \leq 23$, axis ratio $a$:$b$:$c$ = 6:1:1, and rotation period $P\sim8.7$~hr --- within our modelling framework are $\sim20\text{--}40\%$ less discoverable in LSST than spherical objects ($a$:$b$ = 1:1) of comparable size, for slopes $2.5 \leq q_s \leq 4.0$.
This suggests that if a correlation exists between shape and size in the intrinsic ISO population, the discoverability of small ISOs may be more greatly impacted than that of larger ISOs.

Given that ISOs are almost always bright enough for characterisation observations immediately following discovery in LSST, and this holds regardless of whether the ISO is inbound or post-perihelion, it remains reasonable to have ToO programs in reserve at major facilities. 
Immediate diagnosis of any unusual properties will remain a high priority, especially given the sample is likely to be a few tens in number.
However, after initial characterisation, the duration of observability is long enough that many ISOs may be accommodated within regular TAC cycles --- ideally, this would be with multi-semester scheduling continuity.

\subsection{ISOs From the Perspective of a Southern Telescope}

The location of the Simonyi Survey Telescope in the Southern Hemisphere limits the amount of Northern Hemisphere on-sky area included in the survey footprint; in particular, the Solar Apex lies $\sim10^\circ$ outside of the LSST footprint.
Thus, it is expected that the ISOs discovered in LSST will be a biased sampling of the intrinsic \OO{} model.
Our quantification of the discovery biases on ISO discovery in LSST demonstrates the losses incurred from surveying a dominantly Northern-inbound small-body population with a dominantly Southern-pointing telescope.
Orbits discovered in LSST are generally oriented with their perihelia in the Southern Hemisphere (Fig.~\ref{fig:om_Om}).
In contrast to previous expectations that ISOs will be found pre-perihelion \citep{Hoover_2022}, our simulations show that ISOs are equally likely to be discovered pre-perihelion as post-perihelion by LSST (Fig.~\ref{fig:disc_rel}) and are discovered almost uniformly across the LSST footprint (Fig.~\ref{fig:onsky_FDL_allH_q28}).
Even so, ISOs are more likely to be first discovered in LSST near the galactic plane interior to the Solar Circle.
This confirms that most ISOs encounter the Solar System from the direction of the Solar Apex, and suggests that inbound ISOs would be more discoverable in a Northern Hemisphere survey of equivalent limiting magnitude to LSST.

\subsection{Potential Improvements to LSST's ISO Discovery}

The LSST SSP discovery criteria affect the possible ISO discoveries in LSST.
The current moving object discovery requirements are three tracklets on different nights within 15 days.
These criteria result in the discovery of $\sim15 \text{--}40\%$ of all observed objects, depending on the absolute magnitude distribution.
However, additional simulations for $q_s=2.8$ easing the three-tracklet requirement to two tracklets on two nights within the same period increased the discovery completeness of observed ISOs from $\sim34\%$ to $\sim50\%$.
As we find the number of tracklets per object depends on the absolute magnitude distribution of the population ($\S$~\ref{sec:sfd}), this may have a greater impact on an ISO population with a steeper slope~$q_s$.

Realistically, reducing the ISO discovery criteria to require fewer images would decrease the computational accuracy and efficiency.
To first-order, moving objects are discovered by attempting to fit orbits to multiple tracklets and identifying the associated error \citep{Bernstein_2000}.
This is extremely computationally expensive, but recent advances in object detection have shown reductions in this cost.
For example, the LSST SSP will use HelioLinc \citep{Holman_2018} to cluster pre-determined tracklets in a heliocentric frame and link them.
Other methods include a ``shift and stack" approach \citep{Fraser_2024a, Fraser_2024b}.
Both methods have their advantages, but both may be challenged by the low observation count and wide-ranging on-sky motion rate of ISOs.
Undoubtedly, without the constraint of extra observations, moving object detection algorithms will struggle to perform without high false-positive rates.
However, we find that a small number of undiscovered ISOs may be recoverable from simply linking objects across visits in the LSST Deep Drilling Fields, which suggests that there may be additional methods of ISO discovery in the LSST dataset beyond the LSST SSP pipeline.

\subsection{Missions to ISOs}

Since the discovery of \oneI{} and \twoI{}, the possibility of an interstellar object space mission has been raised.
ESA's upcoming \textit{Comet Interceptor} mission is an obvious choice for an ISO flyby.
The mission design of \textit{Comet Interceptor} to wait in the Sun-Earth L2 point until an appropriate target is identified \citep{Jones_2024} is optimal for the uncertainty of when `3I' will pass through the Solar System.
However, as an F-class ESA mission \textit{Comet Interceptor} has a restricted $\Delta v$ budget to manoeuvrer to its target from the L2 position \citep{Sanchez_2021}, so target selection is critical to the mission's success.
Our analysis shows that $\leq 0.95\%$ of the ISOs discovered in LSST will reach a heliocentric distance accessible to \textit{Comet Interceptor}, i.e. have perihelia $q\leq1.35$~au (Fig.~\ref{fig:qiv}).
A mission with a larger $\Delta v$ budget would likely have greater opportunities for the flyby of an ISO.

Recent studies have suggested different strategies for ISO missions.
\citet{Stern_2024} recommended a storage-in-orbit approach with budget $\Delta v \sim 4.0$~km/s.
However, the mission requires up to $\sim40$~days for the Earth and Moon to align to the optimal orientation for an Earth swing-by and the spacecraft to exit the Earth-Moon L1 storage position.
For a similar design, \citet{Landau_2023} estimated that $\sim40\%$ of ISOs within 10~au of the Sun would be accessible for a rendezvous mission.
\citet{Seligman_2018} explored a different strategy: a kinetic impactor producing a plume of internal materials to be analysed, similar to NASA's Double Asteroid Redirection Test \citep{Rivkin_2023} and Deep Impact \citep{A'Hearn_2005} missions.

For all of the mission concepts described above, one of the key drivers for their success is the amount of planning or maneuver time available between an ISO's discovery and its intercept.
We show that even if ISOs are discovered before perihelion ($\sim50\%$ probability; Figure~\ref{fig:disc_rel}), the majority are found by LSST only 60--360~days beforehand (dependent on absolute magnitude slope).
Depending on their trajectory relative to the Earth, most of these may be accessible within the 90-day example flight scenario proposed by \citet{Stern_2024}.

\subsection{Implications for LSST's Cometary ISOs}

While the observability of cometary ISOs in LSST, particularly the selection of a cometary activity model, is left for future work, predictions can be made qualitatively by extrapolation.
Several physical differences between asteroids and comets include their absolute magnitude distribution \citep{Boe_2019}, colours \citep{Jewitt_2015}, phase function \citep{Kiselev_1981, Snodgrass_2011}, and heliocentric brightening \citep{Jewitt_2022}.
Our analysis has investigated the correlations between observability in LSST for all but one of these characteristics.
The last, heliocentric brightening, is due to the evolution of the comet's coma throughout its perihelion passage.
A coma increases the brightness of a comet as it moves closer to the Sun due to the increase in sublimation rate (a function of temperature), scattering of solar light off dust in the coma and the apparent on-sky extent of the coma relative to the observer \citep{Jewitt_2022}.
Comets are anticipated to be discovered in LSST at least 5 years pre-perihelion \citep{Inno_2024}.
Compared to an inactive ISO of equal absolute magnitude, a cometary ISO will be more observable in a flux-limited survey.
To first order, cometary activity provides an equivalent improvement to an object's apparent magnitude as an increased absolute magnitude.
Hence, for a qualitative estimate, the observability results for the absolute magnitude distribution slope $q_s=2.5$ can be considered a potential proxy for a cometary ISO population.
Compared to the steepest slope modelled $q_s=4.0$, the absolute magnitude distribution with $q_s=2.5$ produces a population of ISOs with lower absolute magnitudes (Fig.~\ref{fig:h_dists}),
larger perihelia (Fig.~\ref{fig:qiv}), an equal ratio of prograde and retrograde orbits (Fig.~\ref{fig:qiv}),
larger heliocentric distances at discovery (Fig.~\ref{fig:atDiscovery}),
longer residence times, and a higher likelihood of discovery before perihelion (Fig.~\ref{fig:disc_rel}).
These results qualitatively agree with the comparisons between active and inactive ISOs from \citet{Cook_2016} and \citet{Engelhardt_2017}.

\subsection{Our New Understanding of ISOs in the Era of LSST}

The \OO{} population model predicts complex distributions of velocities, ages and compositions for ISOs in the local neighbourhood.
By applying an absolute magnitude and other physical properties, we have inferred plausible orbital parameter distributions for the ISOs discovered in LSST.

The observed properties of LSST's ISO discoveries will be distinctly biased by the absolute magnitude distribution of the intrinsic galactic population, including orbital parameters (perihelia, eccentricity, inclination, velocity at infinity), absolute magnitudes (median, and $3\sigma$ confidence interval), circumstances at discovery (heliocentric and geocentric distance, on-sky velocity, phase angle, time until perihelion), and observability in LSST (number of observations and tracklets, arc length, characterisation window lengths).
The subtlety here is that the \textit{distributions} of these values for the LSST-discovered \textit{population} will be evidence for the intrinsic absolute magnitude distribution, not their value for any single discovered object.
The resulting debiased absolute magnitude distribution, representing the end sum of all galactic ISO production processes, can then be used to infer the impact of different proposed processes.

Additionally, the number of ISOs discovered over the survey duration depends on the intrinsic absolute magnitude distribution; expectations vary by two orders of magnitude across a factor of $\sim1.5$ in absolute magnitude slope $q_s$.
However, the spatial number density of ISOs is linearly correlated to the number of discoveries.
The result is a degeneracy, where a low absolute magnitude slope and a high number density will produce the same number of LSST discovered ISOs as a high absolute magnitude slope and a low number density.
The distributions listed above must be correctly characterised in the LSST ISO sample to disentangle this degeneracy.
This demonstrates how crucial it is to quantify observational biases to understand the population of ISOs that LSST will observe.

\section{Conclusion}

The LSST presents an unrivalled opportunity to discover a new sample of interstellar objects. 
We summarise our primary findings from a highly realistic simulation of its observations of the \OO\ population model of ISOs:
\begin{itemize}
    \item ISOs are most likely to be found at the edge of observability, due to their single perihelion passage and observable sphere radius. 
    \item Small ISOs observed by LSST are harder to discover than large ISOs.
        The ``limiting absolute magnitude" for ISOs in LSST is $H_r\approx23$, and these objects are about five times less discoverable than the large observed ISOs with $H_r \leq 13$.
    \item The absolute magnitude distribution slope of the intrinsic ISO population can be constrained within $\pm 0.5$ with the discovery of $\sim20\text{--}80$ ISOs. 
    \item The discovery of a single ISO with $H_r\leq10$ will strongly suggest a shallow slope for the intrinsic absolute magnitude distribution. 
            In contrast, the discovery of an ISO similar to \oneI{} ($H_r\sim22.4$) or smaller would not provide any meaningful constraint.
    \item The orbital parameter distributions of discovered ISOs --- perihelia, eccentricity, inclination, and velocity at infinity --- will be indicators of the intrinsic absolute magnitude distribution. 
    \item For absolute magnitude distribution slopes $2.5 \leq q_s \leq 4.0$ and spatial number density estimates from earlier surveys of $n \sim 10^{-1}$~au$^{-3}$, LSST will discover between 6 and 51 ISOs over its ten years. 
    \item ISOs in LSST will have biased orbit orientations ($\omega$ and $\Omega$) and discovery time relative to perihelion due to being discovered by a southern sky survey. 
    ISOs are equally likely to be discovered before as after perihelion.
    \item The twilight microsurvey contributes $<2\%$ of LSST's ISO discoveries for all slopes $q_s$. 
    \item Most ISOs ($>90\%$) will be immediately bright enough for spectroscopic follow-up at their discovery, and of these most will stay above this brightness for about a month. 
    \item Regardless of individual absolute magnitude and the intrinsic absolute magnitude slope, 95\% of ISOs will be bright enough for major ground-based and space telescope follow-up for $\geq2$~months. 
    One in 20 ISOs will be observable for general follow-up for $\gtrsim5$~years if the ISO luminosity function has $q_s=2.8$. 
    \item The majority of ISOs discovered pre-perihelion will be found 60--360 days beforehand, potentially compatible with mission designs for storage-in-orbit with suitable $\Delta v$ budget.
\end{itemize}

Understanding the complex observability and discoverability of ISOs in LSST is vital for interpreting the LSST ISO sample, regardless of whether LSST contains any ISO detections.
The probabilistic description of LSST's ISOs that we quantify offers a path to mitigate the negative impacts 
of ``Hypothesizing After the Results are Known" (or `HARK-ing'), a scientific methodology that retrospectively encourages biases in the direction of future research.
With only two members known so far, the field of small-body Galactic studies is in an ideal position to prepare our models and theories of the little-known ISO population as we move toward the next generation of Solar System surveys.

\begin{acknowledgments}

We thank the LSST Solar System Science Collaboration for manuscript feedback, particularly Sarah Greenstreet, Bryce Bolin, and Dusan Marceta, and the NEO Surveyor Science Workshop and Joe Masiero for helpful discussions and comments.

R.C.D. acknowledges support from the UC Doctoral Scholarship and Canterbury Scholarship administered by the University of Canterbury, a PhD research scholarship through M.T.B.’s Rutherford Discovery Fellowship grant, and LSSTC Enabling Science grant \#2021-31 awarded by LSST Corporation.

M.J.H. acknowledges support from the Science and Technology Facilities Council through grant ST/W507726/1.

M.T.B. and J.C.F. appreciate support by the Rutherford Discovery Fellowships from New Zealand Government funding, administered by the Royal Society Te Ap\={a}rangi.

This work has made use of data from the European Space Agency (ESA) mission
{\it Gaia} (\url{https://www.cosmos.esa.int/gaia}), processed by the {\it Gaia}
Data Processing and Analysis Consortium (DPAC,
\url{https://www.cosmos.esa.int/web/gaia/dpac/consortium}). Funding for the DPAC
has been provided by national institutions, in particular the institutions
participating in the {\it Gaia} Multilateral Agreement.

\end{acknowledgments}

\software{
astropy \citep{AstropyCollaboration_2013,AstropyCollaboration_2018},
numpy \citep{Harris_2020},
pandas \citep{ThepandasdevelopmentTeam_2024},
scipy \citep{Virtanen_2020}
}

\newpage
\appendix

\section{Orbit sampling}\label{sec:appendixOrbSamp}

As detailed in Sec.~\ref{sec:sampling} of this work, \citet{Marceta_2023b} sample the positions of ISOs at a given point in time. 
This is inefficient for simulating non-zero-length surveys, as it requires sampling a much bigger sphere than is actually observable, then propagating the objects along their orbits, many of which will never enter the observable sphere.

Instead, we sample the orbits directly, in terms of pre-encounter velocity \(\mathbf{v}_\infty\), impact parameter \(B\) and angle of impact parameter \(\varphi\).
Note that the distribution of these orbits depends on the length of the survey: longer surveys sample more faster ISOs due to the refresh rate. 
This is due to the residence time \(t_\mathrm{res}(\mathbf{v}_\infty,B,\varphi)\) of objects in the observable sphere becoming negligible compared to the the survey length \(T\); we demonstrate this below.
These five parameters define the orbit; to get the sixth that defines the progress of an object along its orbit, we choose the time of perihelion \(\tau\), relative to the start of the LSST as defined in \texttt{baseline\_v3.3\_10yrs}.
By symmetry, objects on an orbit are distributed uniformly in time; we therefore sample the perihelion time uniformly from \(\tau\in [-\frac{t_\mathrm{res}}{2}, \frac{t_\mathrm{res}}{2} + T]\)
This gives us a sample of ISOs and their orbits that will be in the observable sphere when the LSST is acquiring data.

The flux of ISOs onto an orbit (\(\mathbf{v}_\infty\),\(B\),\(\varphi\)) is given by \citet{Marceta_2023b} Eq. 11:
\[F_{\mathbf{v}_\infty B\varphi}
= n B v_\infty p_{\mathbf{v}_\infty} 
\]
With absolute value of semimajor axis \(A=GM_\odot/v_\infty^2=-a\), perihelion \(q=-A + \sqrt{A^2+B^2}\), and maximum impact parameter for orbits entering the sphere at a given velocity \(B_\mathrm{max}=\sqrt{\rh^2+2A\rh}\), the residence time of object on an orbit (\(v_\infty\), \(B\) \(\varphi\)), the time it spends within sphere of radius \(\rh\) centred on the Sun
is
\begin{equation}\label{eq:tres}
\begin{split}
t_\mathrm{res}(\mathbf{v}_\infty,B)
& = \frac{2}{v_\infty} \int_q^{\rh} \frac{\rh\mathrm{d}\rh}{\sqrt{\rh^2+2A\rh -B^2}}\\
& = \frac{2}{v_\infty} \Bigg( \sqrt{B_\mathrm{max}^2-B^2} - A \log{\bigg(\frac{\sqrt{B_\mathrm{max}^2-B^2}+\rh+A}{\sqrt{A^2+B^2}}\bigg)}\Bigg)
\end{split}
\end{equation}

Therefore the number of objects on an orbit (\(\mathbf{v}_\infty\), \(B\) \(\varphi\)) that will be inside the observable sphere at any point in the survey of length \(T\) is equal to the sum of those in the sphere when the survey begins, plus the number that flow in over time \(T\):
\begin{equation}\label{eq:appendxorbdist}
\begin{split}
\frac{\mathrm{d}N}{\mathrm{d}^3\mathbf{v}_\infty\mathrm{d}B\mathrm{d}\varphi} &= F_{\mathbf{v}_\infty B\varphi} \bigg(t_\mathrm{res}(\mathbf{v}_\infty,B,\varphi) +T\bigg)\\
&= 2n Bp_{\mathbf{v}_\infty}\Bigg( \sqrt{B_\mathrm{max}^2-B^2} - A \log{\bigg(\frac{\sqrt{B_\mathrm{max}^2-B^2}+\rh+A}{\sqrt{A^2+B^2}}\bigg)} +\frac{v_\infty T}{2} \Bigg)
\end{split}
\end{equation}
For \(T=0\), this equals the distribution of orbits of ISOs within a sphere of radius \(\rh\) at a single moment in time, i.e. the distribution of \citet{Marceta_2023b}. 
For \(T\gg t_\mathrm{res}\), the ISOs already in the sphere when the survey is negligible compared to those that have flowed in since, so the distribution is dominated by the refreshing population --- this is the volume-sampling weighted distribution used in \citet{Hopkins_2025} (\(q<\rh\) in Table 1). 
\(\varphi\) is drawn uniformly from 0 to \(2\pi\) independently of all other parameters, so we marginalise over this, and writing \(x=B/B_\mathrm{max}\), \(A' = A/B_\mathrm{max}\) and noting \(\rh+A = \sqrt{A^2+B_\mathrm{max}^2} = B_\mathrm{max}\sqrt{A'^2+1}\) gives
\begin{equation}\label{Bvdist}
\begin{split}
\frac{\mathrm{d}N}{\mathrm{d}^3\mathbf{v}_\infty\mathrm{d}B} = 4\pi n B_\mathrm{max}^2 p_{\mathbf{v}_\infty} \Bigg( x\sqrt{1-x^2} - A'x \log{\bigg(\frac{\sqrt{1-x^2}+\sqrt{A'^2+1}}{\sqrt{A'^2+x^2}}\bigg)} + \frac{v_\infty T}{2B_\mathrm{max}}x \Bigg)
\end{split}
\end{equation}

To sample \(\mathbf{v}_\infty\) and \(B\) we need the cumulative distribution function of B at given values of \(\mathbf{v}_\infty\). 
First we integrate up to a given \(B\):
\begin{equation}\label{eq:Bcdf}
\begin{split}
\int_0^B\frac{\mathrm{d}N}{\mathrm{d}^3\mathbf{v}_\infty\mathrm{d}B}\mathrm{d}B& = 4\pi n B_\textrm{max}^3 p_{\mathbf{v}_\infty} \int_0^x \Bigg(x\sqrt{1-x^2} - A'x \log{\bigg(\frac{\sqrt{1-x^2}+\sqrt{A'^2+1}}{\sqrt{A'^2+x^2}}\bigg)} + \frac{v_\infty T}{2B_\mathrm{max}}x\Bigg)\mathrm{d}x\\
& = 4\pi n B_\textrm{max}^3 p_{\mathbf{v}_\infty}\Bigg(\frac{1-(1-x^2)^{3/2}}{3} - \frac{A'}{2}\Bigg((A'^2+x^2)\log{\bigg(\frac{\sqrt{1-x^2}+\sqrt{A'^2+1}}{\sqrt{A'^2+x^2}}\bigg)}\\
& \quad 
- A'^2\log{\bigg(\frac{1+\sqrt{A'^2+1}}{A'}\bigg)}-\sqrt{A'^2+1}\sqrt{1-x^2}+\sqrt{A'^2+1}\Bigg) + \frac{v_\infty Tx^2}{4B_\mathrm{max}}\Bigg)
\end{split}
\end{equation}
When integrated to \(B=B_\mathrm{max}\), or \(x=1\), this integral is equal to the velocity distribution of orbits passing through the sphere
\begin{equation}\label{eq:vdist}
\begin{split}
\frac{\mathrm{d}N}{\mathrm{d}^3\mathbf{v}_\infty} = 4\pi n B_\textrm{max}^3 \Bigg(\frac{1}{3} - \frac{A'}{2}\Bigg(\sqrt{A'^2+1}
- A'^2\log{\bigg(\frac{1+\sqrt{A'^2+1}}{A'}\bigg)}\Bigg) + \frac{v_\infty T}{4B_\mathrm{max}}\Bigg)\cdot p_{\mathbf{v}_\infty}
\end{split}
\end{equation}
Eq.~\ref{eq:Bcdf} divided by Eq.~\ref{eq:vdist} is equal to the cumulative distribution function of \(B\) given a value of \(\mathbf{v}_\infty\): \(\text{CDF}(B\mid\mathbf{v}_\infty)\).

We can now start sampling! 
Eq.~\ref{eq:vdist} gives the distribution of pre-encounter velocities of the ISOs passing through the sphere. 
Since the ``underlying ISO distribution'' (Table 1 of \citet{Hopkins_2025}) corresponds to \(p_{\mathbf{v}_\infty}\), we reweight those samples by the factor in front of \(p_{\mathbf{v}_\infty}\) in Eq.~\ref{eq:vdist} and resample them to get a sample of \(\mathbf{v}_\infty\).
For each \(\mathbf{v}_\infty\) in this sample, we can evaluate the cumulative distribution \(\text{CDF}(B\mid\mathbf{v}_\infty)\), invert it and draw a value of \(B\). 
Penultimately, for each of these \((\mathbf{v}_\infty,B)\) pairs we draw a perihelion time \(\tau\) from \(\text{Uniform}(-t_\mathrm{res}/2, t_\mathrm{res}/2+T)\) where \(t_\mathrm{res}\) is given by Eq.~\ref{eq:tres}. 
And finally, for each \((\mathbf{v}_\infty,B,\tau)\), we draw a value of \(\varphi\) from \(\text{Uniform}(0,2\pi)\).

\begin{figure}[h]
    \centering
    \includegraphics[width=0.5\linewidth]{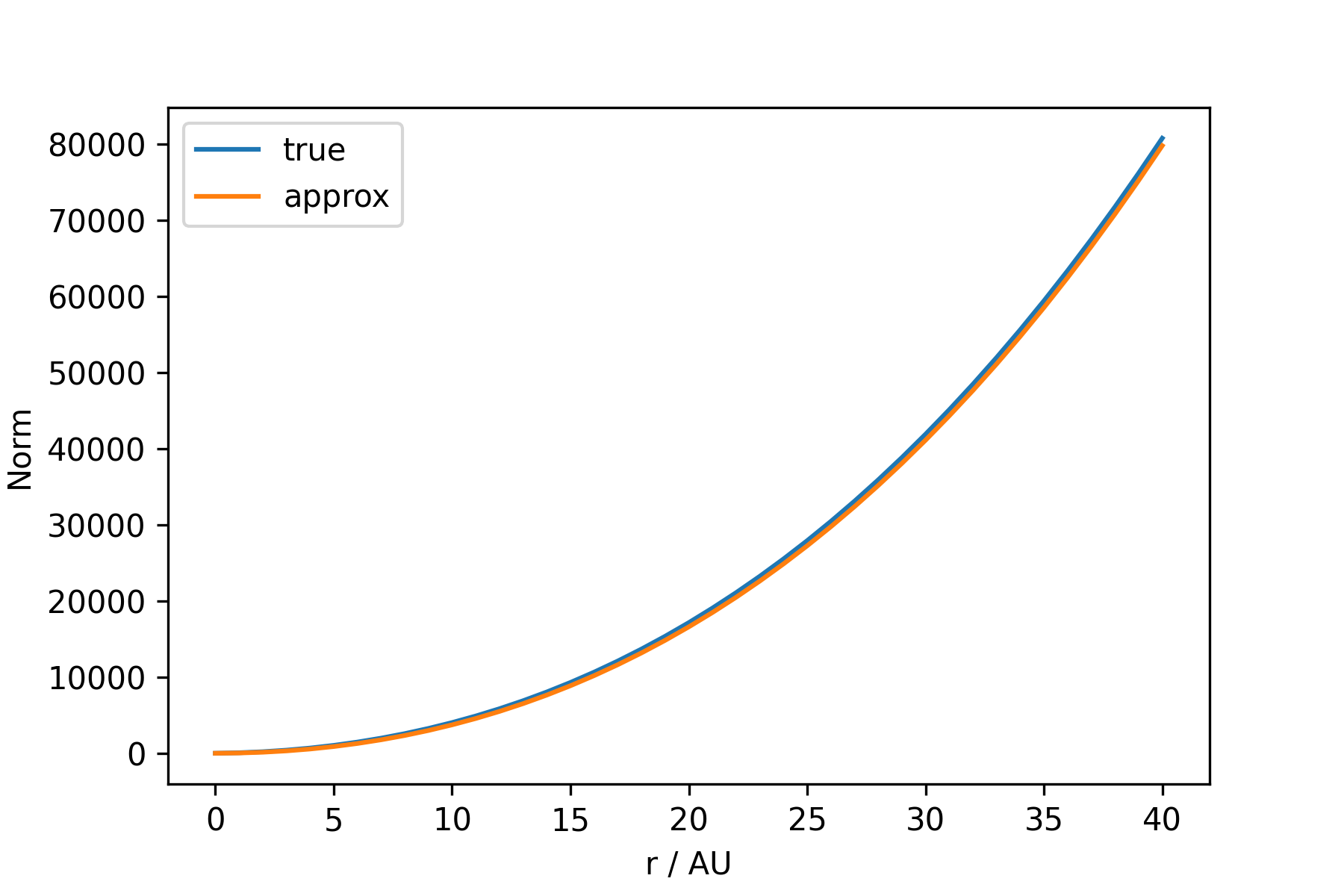}
    \caption{A comparison of the normalisation of the distribution in Eq.~\ref{eq:vdist} to its approximation in Eq.~\ref{eq:normapprox} for different observable spheres of radius \(r\), showing it to be a good approximation.}
    \label{fig:normapprox}
\end{figure}

The integral of Eq.~\ref{eq:vdist} over \(\mathbf{v}_\infty\) is equal to the number of objects entering the observable sphere of radius \(\rh\) over the survey length \(T\).
Since this is difficult to calculate for the non-analytic \OO{} model velocity distribution \(p_{\mathbf{v}_\infty}\), we approximate this with
\begin{equation}\label{eq:normapprox}
    \int{\frac{\mathrm{d}N}{\mathrm{d}^3\mathbf{v}_\infty} \mathrm{d}^3\mathbf{v}_\infty} \approx n\cdot\left(\frac{4}{3}\pi \rh^3 + \pi \rh^2 T(0.029\,\mathrm{au/day})\right)\,.
\end{equation}
A comparison of this approximation to the true value of the integral is plotted in Fig.~\ref{fig:normapprox}, showing it to be very accurate.

\section{The Absolute Magnitude Shallow Slope Effect}\label{appShallowSlope}

\begin{figure}[h]
    \centering
    \includegraphics[width=0.49\linewidth]{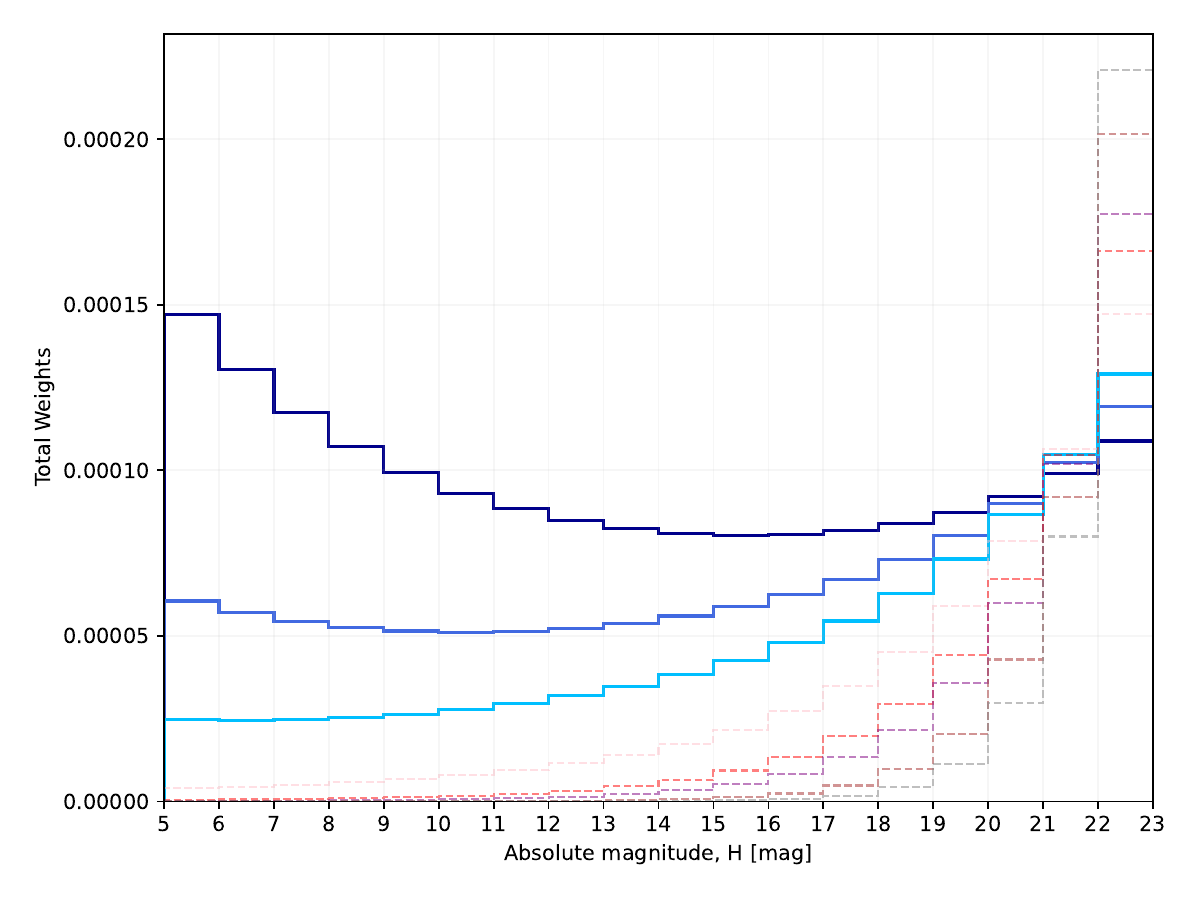}
    \hfill
    \includegraphics[width=0.49\linewidth]{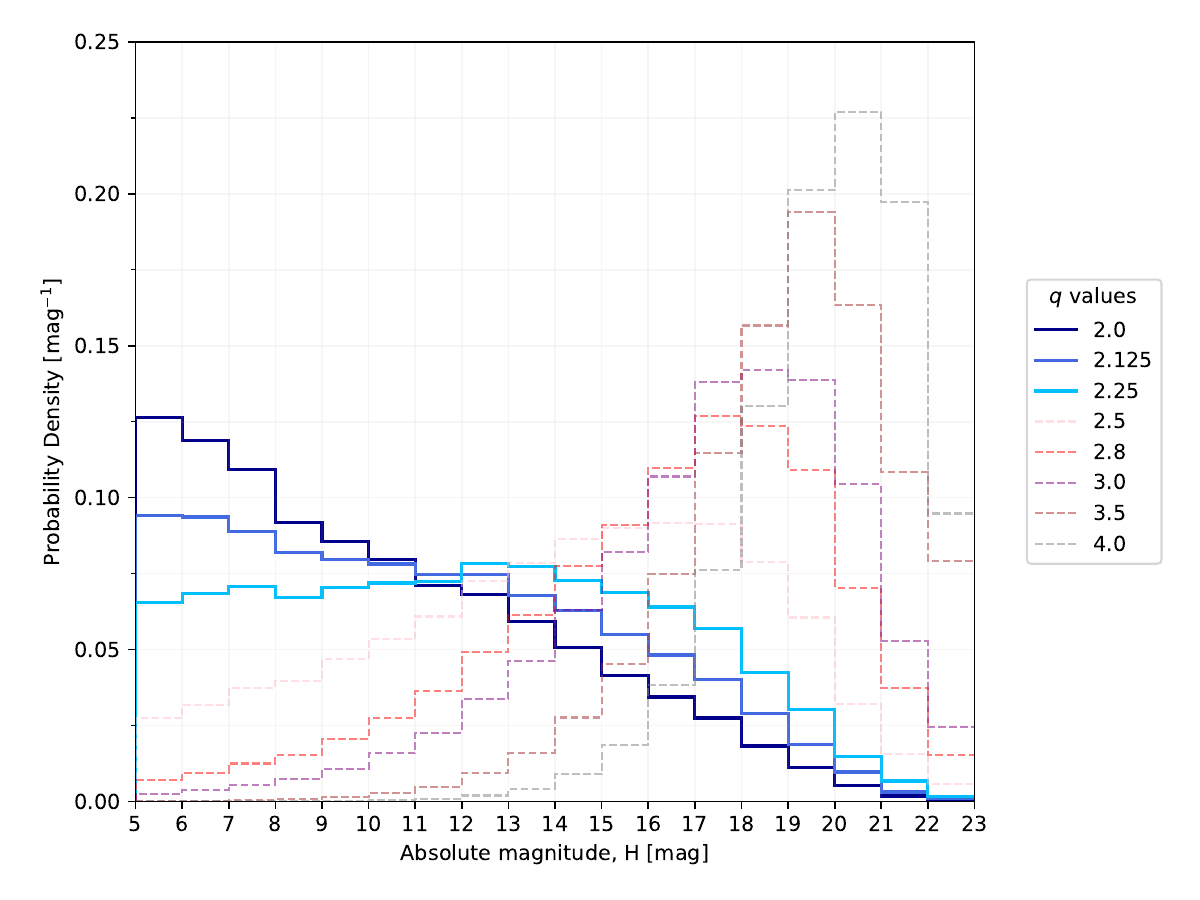}
    \caption{A series of numerical simulations with different absolute magnitude distribution slopes in the range $q_s=[2.0,4.0]$. Left: combined orbital and $H_r$-distribution weightings given by Eq.~\ref{eq:total_weighting}. Right: probability density functions for the discovered ISO population for each slope $q_s$.}
    \label{fig:low_q_phenom}
\end{figure}

ISOs simulated from an absolute magnitude distribution with $q_s=2.0$ ($\alpha = 0.2$) produce an unusual probability density function for the discovered objects in the sample (Fig.~\ref{fig:low_q_phenom}):
more objects with low \(H\) (large diameter) are discovered than objects with high \(H\) (small diameter), despite being less common.
This is due to the ISO population being unbounded in space.
Larger objects have a larger observable sphere, so at some critical slope \(q\), the decreasing frequency of larger objects is counteracted by the increase in the number of objects within the observable sphere.
For size distributions with slopes less than this critical value, the number of objects of a given size contained within the corresponding observable sphere increases with size. 

This effect is unique to the simulation of ISOs.
Unlike any other Solar System small body population, ISOs do not occupy a localised region of the Solar System.
For all small body populations bound to the Solar System, there will always be some heliocentric distance at which the population is complete.
However, ISOs are a Galactic population, so spatial population completeness only occurs across the entire Galaxy.

To constrain the critical slope value, we numerically modelled two additional slope values, $q_s=2.125$ and $q_s=2.25$.
The total weights (Eq.~\ref{eq:total_weighting}) for each slope $q$ value appear approximately parabolic in $H_r$ (Fig.~\ref{fig:low_q_phenom}).
For $q_s\geq2.5$, the minimum weighting always occurs for $5\leq H_r \leq 6$.
As slope $q_s$ decreases, the minimum weighting shifts to larger values of $H_r$ (smaller objects).
The transition occurs around $q_{s, \mathrm{crit}} \approx 2.25 \text{--}2.5$, or $\alpha_{\mathrm{crit}} \approx 0.25\text{--}0.3$.

This can also be determined analytically.
For large objects (small \(H_r\)), as per Eq.~\ref{eq:r_max_obsSphere} objects of absolute magnitude $H_r$ are visible out to a distance of approximately \(r \approx 10^{(m_\mathrm{lim}-H_r)/10}\)~au.
For these objects, the refreshing population is negligible so the observable volume is well approximated by a sphere with volume \(\propto r^3\).
For a power law absolute magnitude distribution \(dN/dH \propto 10^{\alpha H_r}\), the number of visible objects of magnitude $H$ is proportional to \(r^3 10^{\alpha H_r} = 10^{3m_\mathrm{lim}/10}\cdot 10^{(\alpha  -3/10)H_r} \).
Thus there is a critical value \(\alpha_{\mathrm{crit}}=0.3\), or $q_{s,\mathrm{crit}}=2.5$.
Below this value, the number of visible objects increases indefinitely as $H$ decreases.

Both analyses indicate that single-slope absolute magnitude distributions with $q_s<2.5$ represent an improbable model for ISOs.
This is because these absolute magnitude distributions would produce a distribution of detected ISOs that increases indefinitely with decreasing \(H_r\), thus we would expect to discover more large ISOs than small ISOs.
Instead \oneI{}, the one known asteroidal ISO, had an absolute magnitude of \(H_r\sim22.4\), placing it among the smallest ISOs we expect to discover. 
Hence, we consider $q_s=2.5$ the most physically reasonable lower limit for the absolute magnitude distribution slope and use it accordingly for the rest of this work.

\section{Expected Numbers of ISOs}
\label{sec:appendixExpectedNumbers}

Here we compute the number of ISOs that will be discovered in LSST, using the observational and discovery efficiencies of ISOs in LSST determined for different absolute magnitude distributions.

First, we calculate the total observable volume LSST will survey during its 10-year operation.
For an object of absolute magnitude $H_r=5$ and assumed $5\sigma$ limiting $r$-band magnitude $m_5=25$, the corresponding observable sphere has a radius given by Eq.~\ref{eq:r_max_obsSphere}:
\begin{equation}
    r_{H_{\mathrm{min}}}\left(H_{\mathrm{min}}=5\right) = \frac{1}{2}\left(1+\sqrt{1+4\times10^{\frac{25 - 5}{5}}} \right)~\mathrm{au} = 100.5~\mathrm{au}
\end{equation}
The volume of space surveyed by LSST over its duration $T=3652.42$~days is therefore:
\begin{equation}
    V = \frac{4}{3}\pi \left(100.5\right)^3 + \pi \left(100.5\right)^2 \left(3652.42\times0.029\right) = 7.61 \times 10^6~\mathrm{au}^3
\end{equation}
where $0.029$ is the average velocity of ISOs relative to the Sun in au/d.
The number of ISOs in this volume with absolute magnitude $H_r$ is:
\begin{equation}
    N_{\mathrm{sph}} = n V = n \left(7.61 \times 10^6\right)
\end{equation}
where $n$ is the number density for objects with $H_r$ in au$^{-3}$.

The number of ISOs in the $\Delta H$ interval $H_r = \left[ H_{\mathrm{min}}, H_{\mathrm{max}} \right]$, where $H_{\mathrm{max}} = H_{\mathrm{min}} + 1$ of a single slope power law absolute magnitude distribution is given by:
\begin{equation}
    N_{\Delta H} \propto 10^{\alpha \left(H_{\mathrm{min}}+1\right)} - 10^{\alpha H_{\mathrm{min}}}
\end{equation}
where $\alpha$ is defined in relation to the slope parameter $q_s$ by $q_s = 5\alpha + 1$.
The number of ISOs across all $H_r$ intervals is normalised such that $N_{H=22}$ is equal to $N_{\mathrm{sph}}$ for a number density defined with respect to objects with absolute magnitude $H_r=22$:
\begin{equation}
    \begin{split}
         N_{\Delta H} &= \left[10^{\alpha \left(H_{\mathrm{min}}+1\right)} - 10^{\alpha H_{\mathrm{min}}} \right] \frac{N_{\mathrm{sph}}}{10^{22\alpha}} \\
         &= n \left( 7.61\times10^6 \right) \left[ \frac{10^{\alpha \left(H_{\mathrm{min}}+1\right)} - 10^{\alpha H_{\mathrm{min}}}} {10^{22\alpha}}\right]
    \end{split}
\end{equation}

Next, the orbital sampling weightings for each $\Delta H$ interval are calculated by dividing the observable sphere volume at the given $H_{\mathrm{min}}$ value by the largest observable sphere volume:
\begin{equation}
    \begin{split}
        w &= \frac{ \frac{4}{3}\pi r_{H_{\mathrm{min}}}^{3} + \pi r_{H_min}^{2}\left(0.029 T\right) }{\frac{4}{3}\pi 100.5^{3} + \pi 100.5^{2}\left(0.029 T\right)} = \frac{ \frac{4}{3}\pi r_{H_{\mathrm{min}}}^{3} + 105.9\pi r_{H_{\mathrm{min}}}^{2} }{7.61\times10^6}
    \end{split}
\end{equation}
This gives the fraction of objects that enter the ISO population's observable sphere and reach their $H_r$-dependent observable sphere to be potentially observed in LSST.
For example, the interval $H_r=\left[22,23\right]$ where $r_{H_{\mathrm{min}}}=r_{H_{\mathrm{min}}}\left(H_{\mathrm{min}}=22\right)$, has weighting $w=2.95\times10^{-4}$.
This means that if there are $\left(2.95\times10^{-4}\right)^{-1} \approx 3400$ ISOs in the population observable sphere, only one object reaches the $H_r$-dependent observable sphere.
Therefore, the number of ISOs of absolute magnitude $H_r$ which reach their $H_r$-dependent observable sphere is:
\begin{equation}
    \begin{split}
        N_{observable,\Delta H} &= N_{\Delta H} \times w \\
        &= n \left[ \frac{10^{\alpha \left( H_{\mathrm{min}}+1 \right)} - 10^{\alpha H_{\mathrm{min}}}}{10^{22\alpha}} \right] \left( \frac{4}{3}\pi r_{H_{\mathrm{min}}}^{3} + 105.9\pi r_{H_{\mathrm{min}}}^{2} \right)
    \end{split}
\end{equation}

This is then multiplied by the fraction of ISOs $f_{\Delta H}$ in the relevant simulation which is discovered in our LSST survey simulation to get the number of ISOs in the $\Delta H$ interval that are discovered during LSST:

\begin{equation}
    \begin{split}
        N_{discovered,\Delta H} &= N_{observable,\Delta H} \times f_{\Delta H} \\
        &= f_{\Delta H} \times n \left[ \frac{10^{\alpha \left(H_{\mathrm{min}}+1\right)} - 10^{\alpha H_{\mathrm{min}}}}{10^{22\alpha}} \right] \left( \frac{4}{3}\pi r_{H_{\mathrm{min}}}^{3} + 105.9\pi r_{H_{\mathrm{min}}}^{2} \right) \\
    \end{split}
\end{equation}

The discoveries across all $H_r$ intervals are summed to give the total discovered ISOs:
\begin{equation}
    \begin{split}
        N_{discovered,total} &= \sum\nolimits_{H_{\mathrm{min}}=5}^{22}~N_{discovered,\Delta H} \\
        &= n \sum\nolimits_{H_{\mathrm{min}}=5}^{22}~f_{\Delta H} \left[ \frac{10^{\alpha \left(H_{\mathrm{max}}\right)} - 10^{\alpha H_{\mathrm{min}}}}{10^{22\alpha}} \right] \left( \frac{4}{3}\pi r_{H_{\mathrm{min}}}^{3} + 105.9\pi r_{H_{\mathrm{min}}}^{2} \right)
    \end{split}
\end{equation}

We obtain the following prediction for our absolute magnitude distributions:
\begin{equation}
    N_{discovered,total} =
    \begin{cases}
        514n, & q_s=2.5 \\
        244n, & q_s=2.8 \\
        168n, & q_s=3.0 \\
        87.1n, & q_s=3.5 \\
        58.6n, & q_s=4.0
   \end{cases}
\end{equation}

For number density values of $n=0.1$~au$^{-3}$ \citep{Do_2018} and $n=0.24$~au$^{-3}$ \citep{Meech_2017} for $H_r\leq22$ (both determined after the discovery of \oneI{}), our prediction is 6--51 and 14--123 ISOs respectively.
Therefore, we can expect LSST to discover $\sim 10^0 \text{--}10^2$ ISOs during its 10-year nominal survey.

\bibliography{export-bibtex, additional}{}
\bibliographystyle{aasjournal}

\end{document}